\PassOptionsToPackage{unicode=true}{hyperref} 
\PassOptionsToPackage{hyphens}{url}
\documentclass[astrosymb, twocolumn, tighten]{aastex63}
\usepackage{lmodern}
\usepackage{amssymb,amsmath}
\usepackage{ifxetex,ifluatex}
\usepackage{fixltx2e} 
\ifnum 0\ifxetex 1\fi\ifluatex 1\fi=0 
  \usepackage[T1]{fontenc}
  \usepackage[utf8]{inputenc}
  \usepackage{textcomp} 
\else 
  \usepackage{unicode-math}
  \defaultfontfeatures{Ligatures=TeX,Scale=MatchLowercase}
\fi
\IfFileExists{upquote.sty}{\usepackage{upquote}}{}
\IfFileExists{microtype.sty}{%
\usepackage[]{microtype}
\UseMicrotypeSet[protrusion]{basicmath} 
}{}
\IfFileExists{parskip.sty}{%
\usepackage{parskip}
}{
\setlength{\parindent}{0pt}
\setlength{\parskip}{6pt plus 2pt minus 1pt}
}
\usepackage{hyperref}
\hypersetup{
            pdfborder={0 0 0},
            breaklinks=true}
\usepackage{graphicx}
\makeatletter
\def\maxwidth{\ifdim\Gin@nat@width>\linewidth\linewidth\else\Gin@nat@width\fi}
\def\maxheight{\ifdim\Gin@nat@height>\textheight\textheight\else\Gin@nat@height\fi}
\makeatother
\setkeys{Gin}{width=\maxwidth,height=\maxheight,keepaspectratio}
\providecommand{\tightlist}{%
  \setlength{\itemsep}{0pt}\setlength{\parskip}{0pt}}
\ifx\paragraph\undefined\else
\let\oldparagraph\paragraph
\renewcommand{\paragraph}[1]{\oldparagraph{#1}\mbox{}}
\fi
\ifx\subparagraph\undefined\else
\let\oldsubparagraph\subparagraph
\renewcommand{\subparagraph}[1]{\oldsubparagraph{#1}\mbox{}}
\fi

\makeatletter
\def\fps@figure{htbp}
\makeatother

\usepackage[capitalize]{cleveref}
\usepackage{CJKutf8}

\newcommand{\numax}{\ensuremath{{\nu_\text{max}}}}
\newcommand{\chinesename}{{\begin{CJK}{UTF8}{gbsn}(王加冕)\end{CJK}}}

\usepackage{mathptmx,txfonts,tikz}
\usepackage[]{natbib}
\date{}

\begin{document}

\title{Semi-analytic expressions for the isolation and coupling of mixed modes}
\correspondingauthor{Joel Ong}
\email{joel.ong@yale.edu}
\author[0000-0001-7664-648X]{J. M. Joel Ong \chinesename}
\affiliation{Department of Astronomy, Yale University, 52 Hillhouse Ave., New Haven, CT 06511, USA}
 \author[0000-0002-6163-3472]{Sarbani Basu}
\affiliation{Department of Astronomy, Yale University, 52 Hillhouse Ave., New Haven, CT 06511, USA}
\date{June 23, 2020}
\received{May 11, 2020}
\revised{June 8, 2020}
\accepted{June 23, 2020}
\submitjournal{\apj}


\begin{abstract}
In the oscillation spectra of giant stars, nonradial modes may be seen to undergo avoided crossings, which produce a characteristic "mode bumping" of the otherwise uniform asymptotic p- and g-mode patterns in their respective echelle diagrams. Avoided crossings evolve very quickly relative to typical observational errors, and are therefore extremely useful in determining precise ages of stars, particularly in subgiants. This phenomenon is caused by coupling between modes in the p- and g-mode cavities that are near resonance with each other. Most theoretical analyses of the coupling between these mode cavities rely on the JWKB approach, which is strictly speaking inapplicable for the low-order g-modes observed in subgiants, or the low-order p-modes seen in very evolved red giants. We present both a nonasymptotic prescription for isolating the two mode cavities, as well as a perturbative (and also nonasymptotic) description of the coupling between them, which we show to hold good for the low-order g- and p-modes in these physical situations. Finally, we discuss how these results may be applied to modelling subgiant stars and determining their global properties from oscillation frequencies. We also make our code for all of these computations publicly available.
\end{abstract}

\hypertarget{introduction}{%
\section{Introduction}\label{introduction}}

In a strictly ideal sense, stellar oscillations come in two flavours:
acoustic (i.e.~``pressure modes'', or p-modes, deriving their restoring
force primarily from pressure), and buoyant (i.e.~``gravity modes'', or
g-modes, deriving their restoring force primarily from buoyancy). In
solar-like stars, these propagate in mode cavities that are well
separated \citep{unno_nonradial_1989, aertsbook, basu_book_2017}.

The simplest analytic approaches for constructing the frequency
eigenvalues of p- and g-mode oscillations rely on the
Jeffreys-Wentzel-Kramers-Brillouin (JWKB) approximation \citep[see
e.g.~][]{gough_elementary_2007}, and therefore hold good in the limit of
modes of high radial order. The eigenvalues of high-frequency p-modes
follow the approximate asymptotic relation \begin{equation}
    \nu_{nlm} \sim {\Delta\nu} \left(n_p + {l \over 2} + \epsilon_{nlm, p} \right),\label{eq:asp}
\end{equation} where in the limit of high \(n_p\) the phase lag
\(\epsilon_p\) becomes essentially constant with frequency. Likewise,
the frequencies of low-frequency g-modes follow the asymptotic
expression \begin{equation}
    {1 \over \nu_n} \sim \Delta\Pi_l \left(n_g + \epsilon_{nlm, g}\right),\label{eq:asg}
\end{equation} mirroring the standard expression for p-modes. At high
\(n_g\), \(\epsilon_g\) can again be taken to be essentially constant
with frequency, and it is the period spacing \(\Delta\Pi\), rather than
any of the individual g-mode frequencies, which may be used as a
structural or evolutionary constraint
\citep[e.g.~][]{bedding_gravity_2011}.

In evolved solar-like oscillators, these two mode cavities couple
evanescently, leading to mixed modes with both p-like and g-like
character in different parts of the star \citep{osaki_nonradial_1975}.
For bookkeeping purposes, these can be understood as combinations of
fictitious modes of purely g-like and purely p-like character, which are
respectively referred to as ``\(\gamma\)-modes'' and ``\(\pi\)-modes''
\citep{aizenman_avoided_1977, bedding_replicated_2012}. Modes of such
mixed character have been used as sensitive interior probes of the
structure of these stars. Conventional methods of doing so, however,
rely on the accuracy of asymptotic expressions returned from JWKB
analysis. A simplified approach to such analysis returns an approximate
radial dispersion relation \begin{equation}
    k_r^2 \sim -{\omega^2 \over c_s^2}\left(1 - {N^2 \over \omega^2}\right)\left({S_l^2 \over \omega^2} - 1\right),\label{eq:dispersion}
\end{equation} (here \(\Lambda^2 = l(l+1)\), \(N^2\) is the square of
the Brunt-Väisälä frequency, and \(S_l^2 = {\Lambda^2 c_s^2 / r^2}\)
that of the Lamb frequency) such that the wavefunctions are locally
highly oscillatory in regions where \(\omega^2\) is significantly
greater than (respectively, less than) both \(N^2\) and \(S_l^2\) for
p-modes (respectively, g-modes), and decay rapidly otherwise, so that
the JWKB approximation holds good. Correspondingly, the local dispersion
reduces to \(k_r^2 \sim {\omega^2 / c_s^2}\) for acoustic waves
(respectively, \(k_r^2 \sim {S_l^2 N^2 / \omega^2}\) for buoyancy
waves), which may be interpreted as the coefficient of the eigenvalue
term of a corresponding Sturm-Liouville problem.

These naive limits suffice for the study of p- or g-modes in isolation,
in the separate limit of high \(n_g\) or \(n_p\). However, in actual
stars with solar-like convective stochastic mode excitation, only
surface acoustic oscillations at frequencies near \(\numax\), the
frequency of maximum oscillation power, can be measured. We point out
four distinct observational and asymptotic regimes associated with
\(\numax\):

\begin{itemize}
\tightlist
\item
  High \(n_g\) and high \(n_p\), which permits the use of JWKB results
  in both the g- and p- cavities. This is commonly assumed to be the
  case for first-ascent red giant branch stars of intermediate age,
  where many \(\gamma\) modes couple to a single \(\pi\) mode. A
  formidable body of work
  \citep[e.g.~][]{goupil_seismic_2013, deheuvels_seismic_2015, mosser_period_2015, takata_asymptotic_2016, pincon_probing_2020}
  has been assembled based on JWKB expressions for the coupling
  strengths between the two mode cavities, related to this physical
  scenario.
\item
  Low \(n_g\) and low \(n_p\), which precludes the use of JWKB analysis
  altogether;
\item
  Low \(n_p\) and high \(n_g\), which is typical of very evolved
  red-giant stars; in these cases the period spacings of \cref{eq:asg}
  are commonly used for evolutionary constraints \citep[as
  in][]{bedding_gravity_2011}, although the p-modes deviate
  significantly from the asymptotic relation
  \citep{stello_nonradial_2014}.
\item
  Low \(n_g\) and high \(n_p\), which is typical of mixed modes seen in
  subgiants, particularly in the TESS field.
\end{itemize}

These latter two scenarios are characterised by many-to-one coupling
between a sparse set of modes in one mode cavity, and a dense set of
modes in the other. In the case of subgiants, only a few
\(\gamma\)-modes (typically the highest in frequency) couple to the
relatively denser set of \(\pi\)-mode oscillations that subsist in the
convective exterior of a star. As the star evolves, the frequency of the
lowest-order \(\gamma\)-mode increases rapidly relative to those of the
\(\pi\) modes. Evanescent transmission of wave propagation between the
two cavities causes the emergence of the ``avoided crossing''
phenomenon, where the frequencies of the mixed modes are shifted
relative to their uncoupled values, changing smoothly as the star
evolves in such a way as to preserve the ordering of the complete set of
eigenvalues in a continuous fashion throughout this evolution. In
evolved red giants the converse is true; the lowest-order \(\pi\) modes
couple to a dense set of \(\gamma\) modes. Again, as the star evolves,
the frequency of the lowest-order \(\pi\) mode decreases relative to
those of the coupled \(\gamma\) modes.

Since most of the existing theoretical formalism pertaining to the
coupling between \(\pi\) and \(\gamma\) modes relies on JWKB results, it
is not strictly applicable to these latter two scenarios. However, both
of these are of considerable scientific interest. Very evolved red giant
stars (particularly near the tip of the red giant branch) serve as
standard candles and anchor points for isochrone fitting of stellar
populations \citep[e.g.~][]{lee_trgb_1993}. Moreover, avoided crossings
place very strong, albeit model-dependent, seismic constraints on the
structure, ages, and fundamental parameters of subgiants
\citep[e.g.~][]{metcalfe_precise_2010, deheuvels_constraints_2011},
which in turn have been used as benchmarks for comparison between
different measurement techniques
\citep[e.g.~][]{stokholm_subgiant_2019}. Subgiants in particular
dominate the TESS short-cadence seismic sample (owing to constraints on
observational cadence), and are expected to be a substantial fraction of
the PLATO sample as well. We therefore seek a description of mode
coupling that can be applied to such low-order modes.

We present a formalism specifically intended for use in the regime where
JWKB analysis cannot be relied upon to describe the sparse set of
eigenvalues. We first describe a construction of isolated \(\pi\) and
\(\gamma\)-mode eigenfunctions appropriate for such evolved stars
(\autoref{isolated-pi-and-gamma-cavities}). Having done so, we then
derive nonasymptotic expressions for the coupling between the two,
generalising the existing body of JWKB results
(\autoref{coupled-mode-cavities}). We pay particular attention to the
phenomenology of subgiants exhibiting avoided crossings. We finish with
a brief discussion of possible applications to stellar modelling against
observed oscillation spectra
(\autoref{applications-to-stellar-modelling}).

\hypertarget{isolated-pi-and-gamma-cavities}{%
\section{\texorpdfstring{Isolated \(\pi\) and \(\gamma\)
cavities}{Isolated \textbackslash pi and \textbackslash gamma cavities}}\label{isolated-pi-and-gamma-cavities}}

\label{isolated-eigensystems}Linear adiabatic oscillations in a
nonrotating star can be expressed as linear combinations of displacement
eigenfunctions \begin{equation}
    \vec{\xi}(r, \theta, \phi, t) = e^{\pm i\omega t}\left(\xi_r(r) \mathbf{Y}_l^m + \xi_h(r)\mathbf{\Psi}_l^m\right)
\end{equation} which emerge as solutions to the system of differential
equations \begin{equation}
    \begin{aligned}
    {1 \over r^2}{\mathrm d \over \mathrm d r} (r^2 \xi_r) - {g \over c_s^2}\xi_r + \left(1 - {S_l^2 \over \omega^2}\right) {P_1 \over \rho c_s^2} &= {\Lambda^2 \over \omega^2}\Phi_1,\\
    {1 \over \rho} {\mathrm d P_1 \over \mathrm d r} + {g \over c_s^2}P_1 + (N^2-\omega^2)\xi_r &= -{\mathrm d \Phi_1 \over \mathrm d r},\\
    {1 \over r^2}{\mathrm d \over \mathrm d r}\left(r^2 {\mathrm d \Phi_1\over \mathrm d r}\right) - \Lambda^2 \Phi_1 = 4\pi G\rho\left({P_1 \over \rho c_s^2} + {N^2 \over g}\xi_r\right)\\
    \xi_h = {1 \over r \omega^2}\left({P_1\over \rho} + \Phi_1\right).
    \end{aligned}\label{eq:osc}
\end{equation} Here \(P_1(r)\) and \(\Phi_1(r)\) are radial functions
describing the Eulerian pressure and gravitational potential
perturbations under separation of variables, and
\(\mathbf{Y}_l^m = Y_l^m \mathbf{e}_r\) and
\(\mathbf{\Psi}_l^m = \nabla Y_l^m\) are the radial and poloidal vector
spherical harmonics.

Subjecting these to appropriate overdetermined boundary conditions
yields solutions at discrete eigenvalues \(\omega\). In the Cowling
approximation, the perturbations to the gravitational potential are
neglected. \citet{unno_nonradial_1989} introduce the auxiliary variables
\begin{equation}
\begin{aligned}
    \tilde\xi &= r^2 \xi_r \exp\left[-\int_0^r \mathrm d r ~{g \over c_s^2}\right] \equiv r^2 \xi_r h_1(r),\\
    \tilde\eta &= {P_1 \over \rho}\exp\left[-\int_0^r\mathrm d r ~ {N^2 \over g}\right] \equiv {P_1 \over \rho} h_2(r)\label{eq:eta}
    \end{aligned}
\end{equation} in terms of which they obtain the linear system
\begin{equation}
\begin{aligned}
    {\mathrm d \over \mathrm d r}\tilde \xi = -{h_1 \over h_2}{r^2 \over c_s^2}\left(1 - {S_l^2 \over \omega^2}\right)\tilde\eta,\\
    {\mathrm d \over \mathrm d r}\tilde \eta = -{h_2 \over h_1}{1 \over r^2}\left(N^2 - \omega^2\right)\tilde\xi.\label{eq:pair}
\end{aligned}
\end{equation} Either of these quantities can be eliminated in favour of
the other to yield second-order differential equations of the form
\begin{equation}
\begin{aligned}
    {\mathrm d^2 \over \mathrm d r^2}\tilde\xi - {\mathrm d \log |\mathcal{P}|\over \mathrm d r} {\mathrm d \over \mathrm d r}\tilde\xi - \mathcal{PQ} \tilde\xi = 0,\\
    {\mathrm d^2 \over \mathrm d r^2}\tilde\eta - {\mathrm d \log |\mathcal{Q}|\over \mathrm d r} {\mathrm d \over \mathrm d r}\tilde\eta - \mathcal{PQ} \tilde\eta = 0\label{eq:pair2}
\end{aligned}
\end{equation} where \(\mathcal{P}\) and \(\mathcal{Q}\) are the
coefficients on the right-hand-sides of \cref{eq:pair}. The term
\(\mathcal{PQ}\) in both equations yields the dispersion relation of
\cref{eq:dispersion}.

Following the convention of \citet{aizenman_avoided_1977}, we refer to
the eigenvalues of modified versions of these oscillation equations,
where terms corresponding to wave propagation in the classical g-mode
(respectively, p-mode) cavities have been suppressed, as \(\pi\)-mode
(respectively, \(\gamma\)-mode) frequencies. We will refer to these
modified equations as ``isolated oscillation equations''. However, the
prescription by which these modifications are to be done is not uniquely
defined. For instance, for the auxiliary dynamical variables above,
differential equations explicitly in Sturm-Liouville form are recovered
in different frequency regimes for each variable
\citep{unno_nonradial_1989}. In particular, strictly acoustic
propagation (i.e.~with Sturm-Liouville eigenvalues proportional to
\(\omega^2\)) is recovered for \(\tilde\xi\) for \(\omega^2 \gg S_l^2\),
and strictly buoyant propagation (i.e.~with Sturm-Liouville eigenvalues
proportional to \(1 / \omega^2\)) for \(\omega^2 \ll S_l^2\).
Conversely, these regimes for \(\tilde\eta\) are recovered for
\(\omega^2 \gg N^2\) and \(\omega^2 \ll N^2\), respectively.

The choice of which terms of \cref{eq:osc} to suppress in order to
obtain \(\pi\) and \(\gamma\) modes amounts to choosing from the above
limits. To isolate \(\pi\) modes, \citet{aizenman_avoided_1977} suppress
the term proportional to \({S_l^2 P_1 / \omega^2 \rho c_s^2}\) in the
first line of \cref{eq:osc} --- this is equivalent to choosing the limit
\(\omega^2 \gg S_l^2\). Likewise, for \(\gamma\) modes, they suppress
the term \(\omega^2 \xi_r\) in the second line of \cref{eq:osc} --- this
is equivalent to taking the limit \(\omega^2 \ll N^2\). These choices
were motivated by a superficial resemblance to Sturm-Liouville
eigenvalue terms proportional to \(1/\omega^2\) and \(\omega^2\),
bearing the interpretations of buoyant and acoustic wave propagation,
respectively.

\begin{figure}[htbp]
    \centering
    \includegraphics[trim=.25cm .75cm .25cm .15cm,clip]{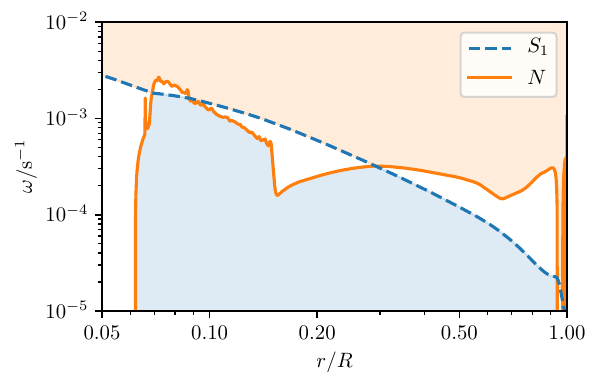}
    \includegraphics[trim=.25cm .25cm .25cm .15cm,clip]{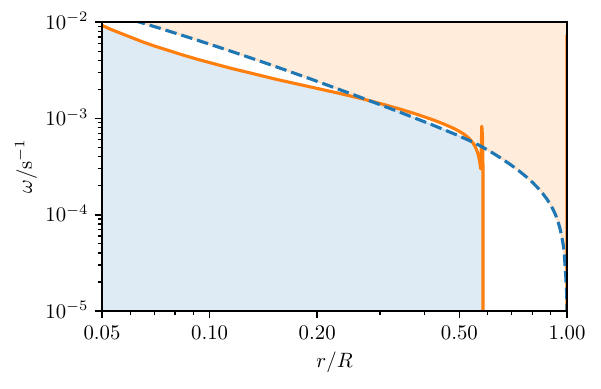}
    \caption{Propagation diagrams for dipole modes with respect to $16M_\Sun$ main-sequence-turnoff (top panel) and $1M_\Sun$ subgiant (bottom panel) MESA evolutionary models. Classical g-modes propagate in the shaded blue regions, and p-modes in the orange regions. \label{fig:prop}}
\end{figure}

These choices yield merely sufficient, but not necessary, conditions for
the propagation of waves of these respective types.
\citet{aizenman_avoided_1977} applied them to study high-mass stars with
convective cores, for which e.g.~the outer boundary of the g-mode cavity
in the limit of low frequency is set by the Lamb frequency (which we
show in the top panel of \cref{fig:prop}). For typical evolutionary
models of lower-mass subgiants and red giants exhibiting solar-like
pulsations, however, the Brunt-Väisälä frequency determines the outer
limit of g-mode propagation, and the Lamb frequency determines the inner
limit of p-mode propagation (bottom panel of \cref{fig:prop}).
Therefore, we claim that this physical scenario requires taking limits
in the converse sense to those taken in \citet{aizenman_avoided_1977}.

Of these two converse choices, \citet{ball_surface_2018} have previously
employed the limit \(\omega^2 \gg N^2\) to perform numerical
computations of \(\pi\)-mode frequencies for red giants in the
low-\(n_p\) (low-frequency acoustic) regime. In principle, this should
be done by suppressing the term proportional to \(N^2 \xi_r\) in the
second line of \cref{eq:osc}. We assert that the complementary limit,
\(\omega^2 \ll S_l^2\), which may be implemented by suppressing the term
proportional to \(P_1 / \rho c_s^2\) in the first line of \cref{eq:osc},
will yield pure-buoyancy \(\gamma\)-mode oscillations even in the
low-\(n_g\) (i.e.~high frequency buoyancy) regime, which is precisely
what is required for subgiant avoided crossings.

Although these choices of which terms of \cref{eq:osc} to suppress are
motivated by asymptotic considerations, in the sense that we identify
terms to suppress that would vanish in the relevant high- or
low-frequency limits, we stress that the isolated systems of equations,
where such terms have been suppressed \emph{a priori}, can be employed
even in frequency regimes that do not satisfy these asymptotic demands.
We will see that this merely results in other terms appearing elsewhere
in our analytic formulation for the coupled system, that vanish when
these asymptotic conditions are satisfied.

\hypertarget{approximate-analytic-formulation}{%
\subsection{Approximate analytic
formulation}\label{approximate-analytic-formulation}}

We first justify our choice of isolation for the \(\gamma\)-mode cavity
as yielding purely buoyant wave propagation. To simplify our analysis,
we begin by examining mode isolation in the Cowling approximation. Since
we intend to study the behaviour of wave propagation in frequency
regimes where standard JWKB methods cannot be applied, we turn instead
to the method of undetermined phases, which returns exact results that
are accurate to the same level of approximation as of the underlying
differential system. The typical scenario where this method is employed
involves a boundary value problem of Schrödinger form, \begin{equation}
    {\mathrm d^2 \over \mathrm d x^2}y + \left(k^2 - V(x)\right)y = 0,\label{eq:schr}
\end{equation} with eigenvalues \(k\) over the domain \([0, X]\).
\(V(x)\) is assumed to be small except near these boundaries, which are
singular points where the solutions \(y\) vanish. We substitute ansatz
solutions for the eigenfunctions
\(u_k(x) \sim A(k, x) \sin(k x - \delta(k, x))\) near \(x = 0\),
demanding that \(u'_k(x) \sim k A(k, x) \cos (k x - \delta(k, x))\).
This yields the constraint on the inner phase function \(\delta(k, x)\)
that \begin{equation}
    {\mathrm d \over \mathrm d x}\delta(k, x) \sim {V(x) \over k}\sin^2(kx - \delta(k, x)).
\end{equation} For \(u_k(x)\) to vanish at the inner boundary,
\(\delta(k, x)\) must vanish at \(x = 0\); this constitutes an initial
value problem (IVP) for \(\delta(k, x)\), which can then be integrated
to any reference point \(x_0\) in \([0, X]\). A similar IVP can be set
up at the outer boundary for a corresponding outer phase function
\(\alpha(k, x)\). These two definitions of the eigenfunctions are
consistent at any given matching point \(x_0\) only if
\(\sin (kx_0 - \delta(k, x_0)) = \sin (k(x_0 - X) - \alpha(k, x_0))\) up
to sign, whence emerges an eigenvalue equation \begin{equation}
    kX + (\alpha(k, x_0) - \delta(k, x_0)) \equiv kX - \pi \epsilon(k) = n\pi,
\end{equation} yielding eigenvalues \(k_n\) that satisfy this expression
for integers \(n\). Once these eigenvalues are known, the eigenfunctions
can then be recovered (up to overall constant factor) by solving a
complementary IVP \begin{equation}
    {\mathrm d \over \mathrm d x}A(k_n, x) = {A(k_n, x)V(x) \over k_n} \sin(k_nx - \delta(k_n, x)) \cos(k_nx - \delta(k_n, x))
\end{equation} from the inner boundary, holding \(k_n\) fixed.

The method of undetermined phases has been employed in the study of
p-mode oscillations
\citep{roxburgh_vorontsov_1996, roxburgh_ratio_2003, ong_structural_2019},
but to our knowledge it is not commonly used to study the g-mode cavity,
since historically \(\Delta\Pi\) has sufficed for most applications of
g-modes as constraints on stellar interiors. Since the method returns
exact results, however, it is ideally suited to working in the low-\(n\)
regime where the JWKB approach is known to fail.

Following the discussion above, we take the radial displacement
functions of \(\gamma\)-modes to satisfy the reduced expression
\citep{unno_nonradial_1989}\begin{equation}
    {\mathrm d^2 \over \mathrm d r^2} \tilde\xi - \left({\mathrm d \over \mathrm d r} \log h\right)\left({\mathrm d \over \mathrm d r} \tilde\xi\right) + {N^2\Lambda^2\over r^2}\left({1 \over \omega^2} - {1 \over N^2}
    \right)\tilde\xi = 0,
\end{equation} where \(h(r) = h_1(r)/h_2(r)\). As discussed previously,
this expression is obtained by suppressing the term \(P_1/\rho c_s^2\)
in the first line of \cref{eq:osc}. From this, we recover an equation of
Sturm-Liouville form \begin{equation}
    {\mathrm d \over \mathrm d r}\left({1 \over h}{\mathrm d \over \mathrm d r}\tilde\xi\right) + {N^2\Lambda^2\over h r^2}\left({1 \over \omega^2} - {1 \over N^2}\right)\tilde\xi = 0 \label{eq:sturmliouville}
\end{equation} with eigenvalues \(1 / \omega^2\).

To put this into Schrödinger form, we choose to change coordinates to
the buoyancy radius \citep{tassoul_asymptotic_1980} \begin{equation}
    f_l = \int_0^r \mathrm d r~{N \Lambda \over r},\label{eq:buoyancy}
\end{equation} which has units of frequency. Moreover, to eliminate the
damping term, we choose a new dynamical variable with an integrating
factor \(\psi = e^u \tilde\xi\), where \begin{equation}
\begin{aligned}
    2u' = -{\mathrm d \over \mathrm d f_l} \log h + {\mathrm d \over \mathrm d f_l} \log {N \over r} \\ \implies \psi = \xi_r \sqrt{h_1h_2 r^3 N} = \xi_r \sqrt{r^3 \rho N}. \label{eq:integratingfactor}
\end{aligned}
\end{equation} After some manipulation \citep[see
e.g.~][]{gough_elementary_2007} this yields an equation of Schrödinger
form \begin{equation}
    \psi'' + \left({1 \over \omega^2} - V_{g, l}(f_l)\right)\psi = 0, \label{eq:potential}
\end{equation} where the buoyancy potential \(V_g\) is given as
\begin{equation}
    V_{g, l}(f_l) = {1 \over N^2}
     + u''(f_l) + (u'(f_l))^2. \label{eq:potential2}
\end{equation} Note that both the buoyancy coordinate and potential
depend on the degree \(l\). Limiting our attention to the
\(\gamma\)-mode cavities found in first-ascent red giants and subgiants
with mixed modes, we additionally observe that as \(r\to0\),
\(N^2 \sim N_0^2 r^2\) for some constant \(N_0^2\). Accordingly,
\(f \sim r\) as \(r\to 0\), and the buoyancy potential is singular at
the central point, which is then a regular singular point of the
differential equation. Likewise, the outer boundary of the
\(\gamma\)-mode cavity is defined by the inner boundary \(r_0\) of the
convection zone, where \(N^2 = 0\) also. This is also a singular point
of the differential equation, which is regular only if the leading order
behaviour of \(N^2\) is either linear or quadratic in \(r - r_0\)
inwards of the boundary. The domain of the associated boundary value
problem is then \(f_l \in [0, F_l]\), where
\(F_l = f_l(r_0) = \int_0^{r_0} (N \Lambda / r) \mathrm d r\). Since
both boundaries are singular points, solutions can be assumed to vanish
there.

With this in hand, we then construct the inner phase function as the
solution to the IVP \begin{equation}
    {\mathrm d \over \mathrm d f_l}\delta_{g,l}(\omega, f_l) = \omega V_{g, l}(f_l) \sin^2\left({f_l \over \omega} - \delta_{g, l}(\omega, f_l)\right), \label{eq:IVP}
\end{equation} and likewise for the outer phase function, following the
above procedure; this yields at last the eigenvalue equation
\begin{equation}
    {F_l \over \omega_{nl}} \sim \pi \left(n + \epsilon_{l, g}(\omega_{nl})\right). \label{eq:eigenvalue}
\end{equation} Comparing this with \cref{eq:asg} yields the usual
asymptotic expression for the period spacing in the regime of constant
\(\epsilon_g\) \citep{tassoul_asymptotic_1980}, \begin{equation}
    \Delta\Pi_l = {2\pi^2 \over F_l}={2\pi^2 \over \sqrt{l(l+1)}} \left(\int_0^{r_0} \mathrm d r {N \over r} \right)^{-1}. \label{eq:dpi}
\end{equation}

\begin{figure}
\centering
\includegraphics{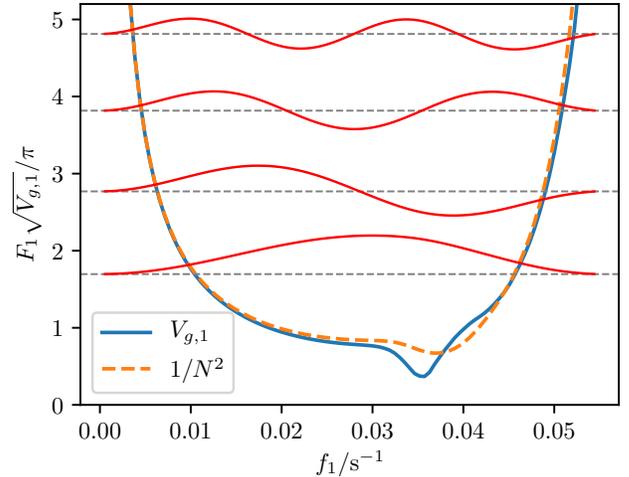}
\caption{\(l=1\) eigenstates of the \(\gamma\)-mode cavity in the
Cowling approximation for a 1\(M_\Sun\) subgiant evolutionary model,
showing both the buoyancy potential and the predominant contribution
from the inverse Brunt-Väisälä frequency.\label{fig:eigenstates}}
\end{figure}

For the purposes of our subsequent discussion and analysis, we consider
evolutionary tracks of stellar models constructed using release 10398 of
the MESA stellar evolution code
\citep{mesa_paper_1, mesa_paper_2, mesa_paper_4}, generated with
solar-like fundamental parameters (i.e.{[}Fe/H{]} = 0, with
solar-calibrated mixing length and helium abundance). We show in
\cref{fig:eigenstates} the Cowling-approximation eigensystem associated
with the \(\gamma\)-mode cavity for a \(1 M_\Sun\) model subgiant
undergoing an avoided crossing (\(\Delta\nu = 72 ~\mu\)Hz), alongside
with the square root of the buoyancy potential, normalised by the
asymptotic period spacing as given by \cref{eq:dpi}. This choice of
scaling places eigenvalues at integer steps of the vertical axis in the
asymptotic regime, as described by \cref{eq:asg}. We see that this is
more or less the case, consistent with the known properties of pure
buoyancy waves. This is the basis of our interpretation of these
\(\gamma\)-modes as yielding purely buoyant wave propagation.

We have also plotted the contribution to this buoyancy potential from
the inverse Brunt-Väisälä frequency, which we see essentially dominates
the dynamics of the system, especially near the core. However, the
remaining terms involving \(u\) are also singular at the outer boundary
for typical subgiant evolutionary models, and cannot be neglected.

\hypertarget{numerical-evaluation-of-gamma-modes}{%
\subsection{\texorpdfstring{Numerical Evaluation of \(\gamma\)
modes}{Numerical Evaluation of \textbackslash gamma modes}}\label{numerical-evaluation-of-gamma-modes}}

The Cowling approximation is known to hold increasingly well at high
order and degree, and we might be doubtful as to its applicability to
subgiant avoided crossings in particular, which are observed at low
order and low degree. Indeed, as seen in \cref{fig:eigenstates}, the use
of the Cowling approximation yields a fictitious \(n=0\) dipole mode,
which implies periodic oscillations of the centre of mass. This is known
to be forbidden under the full system of equations
\citep{jcd_isolated_1976, jcd_dipolar_2001}. Moreover, the
Sturm-Liouville form of \cref{eq:sturmliouville} implies orthogonality
with respect to a different inner product than is obtained for the full
system of equations. We therefore find it prudent to compare our results
to those obtained without the Cowling approximation for the remaining
dipole modes.

We first note that it is in fact possible under some circumstances to
perform a similar analysis without recourse to the Cowling
approximation. For instance, the \(l=1\) oscillation equations admit a
reduction to second order
\citep{takata_asymptotic_2016, pincon_probing_2020} that essentially
preserves the dispersion relation of \cref{eq:dispersion}. However, this
formulation is not applicable to modes of higher degree, and in any case
the displacement eigenfunctions are not directly recovered from this
construction: the dynamical variables there contain the perturbation to
the gravitational potential and its derivative. Eliminating these
requires a second, auxiliary set of differential equations to be solved.
Since we require the eigenfunctions for the computations described later
in \autoref{coupling-strengths}, we choose not to pursue this approach
further.

Instead, we seek explicit recourse to numerical methods here. To this
end, we used the GYRE pulsation code \citep{townsend_gyre_2013} to
compute mixed mode and (with appropriate modifications) \(\gamma\)-mode
frequencies, without using the Cowling approximation, for a series of
subgiant/early red-giant models along a \(1M_\Sun\) MESA evolutionary
track. More details about our modifications to GYRE can be found in
\autoref{implementation-details}.

\begin{figure}
\centering
\includegraphics{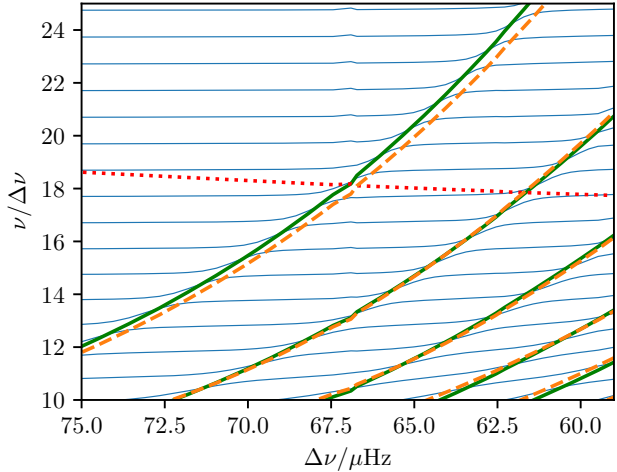}
\caption{Evolution of mixed modes and \(\gamma\) modes for the first few
avoided crossings of a \(1 M_\Sun\) evolutionary track. Mixed modes are
shown with the thin blue lines, with \(\gamma\)-mode frequencies shown
with the solid green line (as computed under the Cowling approximation)
and orange dashed line (as computed from the full system of equations).
The red dotted line shows the evolution of \(\numax\) for the same
models.\label{fig:avoided}}
\end{figure}

We compare results from GYRE with our solutions to the
Cowling-approximation Sturm-Liouville problem for one of these
evolutionary tracks in \cref{fig:avoided}. We see that both solutions
exhibit broadly similar morphology, tracking the implicit
\(\gamma\)-modes traced out by the avoided crossings. For the first
avoided crossing specifically, we also see that both approaches slightly
underestimate the avoided crossing frequency. We will show in
\autoref{coupling-strengths} that the eigenvalues of the \(\gamma\)-mode
system are in general not sufficient to predict the frequency of the
avoided crossing; a first-order correction term must also be computed.
These shortcomings notwithstanding, however, we claim that this further
demonstrates that our interpretation of these \(\gamma\)-modes as purely
buoyant waves is at least qualitatively correct.

\begin{figure}[htbp]
    \centering
    \includegraphics[trim=0 .75cm 0 .15cm,clip]{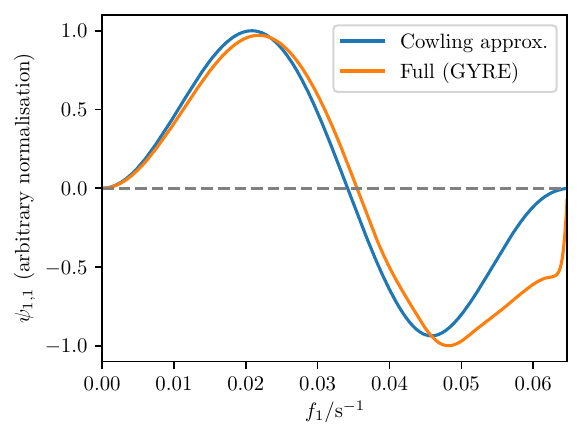}
    \includegraphics[trim=0 0cm 0 .15cm,clip]{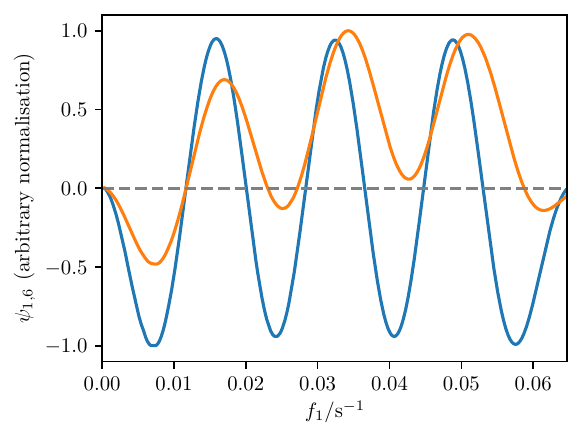}
    \caption{$l=1, n=1$ (top panel) and $n=6$ (bottom) eigenfunctions computed with the Cowling approximation in Sturm-Liouville form, and with the full set of pulsation equations using GYRE, for a 1 $M_\Sun$ subgiant evolutionary model undergoing an avoided crossing. \label{fig:boundary}}
\end{figure}

The other unexpected feature of \cref{fig:avoided} is the increasing
discrepancy with the full-system eigenvalues at low frequencies, where
we would \emph{a priori} expect the Cowling approximation to hold
increasingly well. We attribute this to the singular behaviour of the
outer boundary of the buoyancy cavity in GYRE's numerical scheme. To
illustrate this, we show the \(l=1, n=1\) eigenfunction returned from
our solution in buoyancy coordinates, and from GYRE, in
\cref{fig:boundary}. We see in the top panel that the behaviour of
GYRE's solution is pathological at the outer boundary. This problem
becomes increasingly severe at higher orders; for example, the \(n=6\)
\(\gamma\)-eigenfunction for the same model does not even have the
correct number of zero crossings.

Specifically, the outer boundary condition of the eigenvalue problem is
applied to \(\psi\) at \(f_l = F_l\) in our Cowling-approximation
solution, and to (a dimensionless analogue of) \(\xi_r\) at \(r = R\) in
GYRE. For evolutionary models with outer convection zones, these are
formally inequivalent in the following manner: irrespective of GYRE's
boundary conditions, the corresponding \(\psi\) automatically vanishes
at the endpoints via \cref{eq:integratingfactor} without regard for
regularity (neglecting any radiative atmosphere). However, since the
Brunt-Väisälä frequency vanishes outside of the buoyancy cavity, the
transformation from \(r\) to \(f_l\) defined by \cref{eq:buoyancy} is
formally also degenerate (as the entire convection zone is mapped to a
single point at \(f_l = F_l\)), yielding the discontinuous behaviour
that we see in \cref{fig:boundary}. We verified this by computing
\(\gamma\)-mode eigenfunctions of a polytrope with index \(n=3\), for
which, in the absence of an outer convective zone, we were able to
recover the correct number of nodes even at high radial order.

To recover regularity with respect to the buoyancy coordinate, boundary
conditions expressed in terms of derivatives taken with respect to
\(f_l\), rather than \(r\), must be imposed at the outer boundary. We
note that \begin{equation}
    {\mathrm d \over \mathrm d f_l} = {r \over N} {\mathrm d \over \mathrm d r};
\end{equation} since \(N \to 0\) at the outer boundary, the boundary
conditions there must then involve higher derivatives of the dynamical
variables (with respect to \(r\)) to remain regular via L'Hôpital's
rule. The changes required to do this in GYRE are substantial and beyond
the scope of this work. For the purposes of our subsequent analysis it
suffices merely to note that this is a numerical artifact that is least
severe at low order --- fortuitously, this still permits the study of
the lowest-order avoided crossings.

\hypertarget{numerical-evaluation-of-pi-modes}{%
\subsection{\texorpdfstring{Numerical evaluation of \(\pi\)
modes}{Numerical evaluation of \textbackslash pi modes}}\label{numerical-evaluation-of-pi-modes}}

\begin{figure}
\centering
\includegraphics{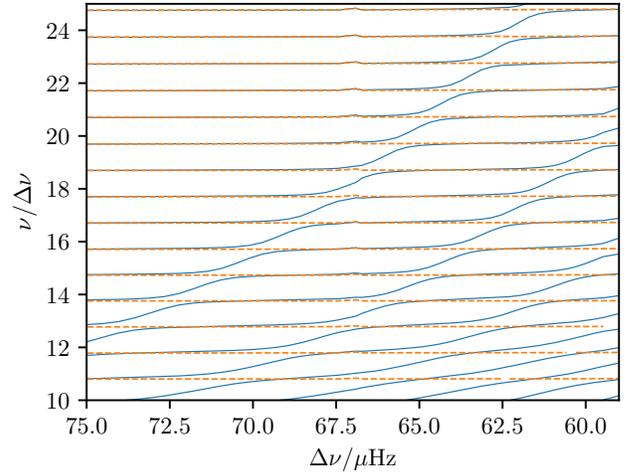}
\caption{Evolution of mixed modes and \(\pi\) modes near the first few
avoided crossings of a \(1 M_\Sun\) evolutionary track. Mixed modes are
shown with the thin blue lines, with first-order corrected \(\pi\)-mode
frequencies shown with the orange dashed lines. \label{fig:pi}}
\end{figure}

To complete this discussion of cavity isolation, we turn our attention
to the \(\pi\) mode cavity. For subgiants undergoing avoided crossings,
the acoustic modes are of relatively high order (\(>10\)), and JWKB
expressions are largely applicable. Moreover, for acoustic modes, the
relevant radial coordinate required to recover an expression of
Sturm-Liouville form is the acoustic radial coordinate, \begin{equation}
    t(r) = \int_0^r {\mathrm d r \over c_s}, \label{eq:acousticradius}
\end{equation} which is well-behaved everywhere in the interior of the
star. Although our ability to accurately predict p-mode frequencies is
affected by the surface term, we expect it to afflict both the full
mixed modes and \(\pi\)-modes in much the same manner, so long as the
Brunt-Väisälä frequency at the surface is unchanged (for more details,
see \autoref{implementation-details}). For the sake of demonstration, we
once again compare the evolution of \(\pi\) modes in our formulation
with the mixed modes returned from an evolutionary track in
\cref{fig:pi}. We see that these \(\pi\)-modes adhere to the p-mode
asymptotic relation \cref{eq:asp} even where the eigenvalues of the
coupled system undergo avoided crossings; we interpret these as being
purely acoustic waves, in agreement with \citet{ball_surface_2018}.

\hypertarget{coupled-mode-cavities}{%
\section{Coupled mode cavities}\label{coupled-mode-cavities}}

\label{coupling-strengths}We have demonstrated that our choices of
isolated mode cavities can be meaningfully interpreted as separately
supporting purely buoyant and purely acoustic waves. As seen in
\cref{fig:avoided,fig:pi}, the eigenvalues of the coupled system exhibit
an ``avoided crossing'' phenomenon over the course of stellar evolution.
Expressions for the frequencies of such avoided crossings are
generically derived by way of a mechanical analogy with a coupled system
of harmonic oscillators
\citep[e.g.~][]{deheuvels_insights_2010, benomar_masses_2012}. The
standard analytic approach here is to find the eigenvalues of some real,
symmetric matrix \begin{equation}
    \mathbf{L} = \mathbf{H}_0 + \alpha \mathbf{V}, \label{eq:perturbation}
\end{equation} where \(\mathbf{H}_0\) is diagonal and nearly degenerate;
the introduction of the coupling matrix \(\mathbf{V}\), with
off-diagonal elements, lifts this degeneracy. Once in this form, the
avoided crossing can be shown to emerge, e.g.~by application of
perturbation theory \citep{vonneumann_avoided_1929}. The basis of this
analogy is such that the matrix \(\mathbf{L}\) describes the time
evolution of some set of dynamical quantities \(\mathbf{y}\) of these
model coupled oscillators as \begin{equation}
    {\mathrm d^2 \over \mathrm d t^2} \mathbf{y} = \mathbf{Ly}.
\end{equation} As far as subgiant avoided crossings are concerned, this
matrix is ordinarily assumed \emph{a priori} to be of some ansatz
parametric form with constant coupling between the \(\pi\) and
\(\gamma\) cavities, motivated by the \(2\times2\) case; to our
knowledge, no explicit construction exists that relates it to properties
of stellar structure. We attempt such a construction in this section.

The perturbed momentum equation for a single mode \(\xi_i\) with
time-dependent coefficient \(c_i\) permits the construction of a
time-dependent operator equation over the Hilbert space of vector
displacement eigenfunctions
\citep{eisenfeld_completeness_1969, jcd_hilbert_1981} in the form
\begin{equation}
    {\mathrm d^2 \over \mathrm d t^2} c_i \xi_i \equiv \hat{\mathcal{L}} c_i \xi_i = -\omega_i^2 c_i\xi_i
\end{equation} where the time dependence is carried entirely by the
coefficient \(c_i\), which in turn is a function only of time (this is
equivalent to working in the Schrödinger picture in quantum mechanics).
For a general state in this Hilbert space, expressed as a linear
combination of eigenfunctions \(\vec{\xi} = \sum_j c_j \vec{\xi}_j\),
the corresponding evolution goes as \begin{equation}
    \sum_j \left({\mathrm d^2 \over \mathrm d t^2}c_j\right) \vec{\xi}_j = \hat{\mathcal{L}} \sum_j c_j \vec{\xi}_j = - \sum_j c_j \omega_j^2 \vec{\xi}_j, \label{eq:evol}
\end{equation} where this linear operator acts independently on each of
its eigenfunctions \(\vec{\xi}_j\), which emerge as solutions to
\cref{eq:osc}. In the case where the \(\xi_i\) form a complete
orthogonal basis, we can recover each of these time-dependent
coefficients by taking inner products under the choice of normalisation
such that \begin{equation}
    \left<\xi_i, \xi_j\right> = \int \mathrm d^3 x~\rho \vec{\xi}_i\cdot \vec{\xi}_j \equiv \delta_{ij}.\label{eq:orthonormal}
\end{equation} Put differently, \cref{eq:osc} is a time-independent
problem that yields the eigensystem of the Hermitian
integro-differential operator \(\hat{\mathcal{L}}\) (which provides some
natural orthonormal basis on the Hilbert space by the spectral theorem).
By contrast, \cref{eq:evol}, which is time-dependent, instead relates
the time evolution of vectors in the Hilbert space to the action of this
operator.

Let us now consider the time evolution of \(\gamma\) and \(\pi\)-mode
eigenstates, which are not eigenstates of \(\hat{\mathcal{L}}\).
Instead, they are solutions to modified versions of \cref{eq:osc}, where
different terms have been suppressed to isolate the mode cavities. In
this abstract operator notation, we consider the \(\pi\) modes to be the
eigenstates of the operator \(\hat{\mathcal{L}}_\pi\), representing the
modified momentum equation, and the \(\gamma\) modes to be those of a
different operator \(\hat{\mathcal{L}}_\gamma\). To use \cref{eq:evol},
we relate these modified operators to the the original set of equations
as the sum of the modified (e.g.~\(\pi\)-mode) operator and some
``remainder'' operator: \begin{equation}
    \hat{\mathcal{L}} = \hat{\mathcal{L}}_\pi + \hat{\mathcal{R}}_\pi,
\end{equation} where this remainder operator is simply the term that has
been suppressed in order to yield the isolated system of equations for
\(\pi\)-modes. In this case, we can easily see that
\(\hat{\mathcal{R}}_\pi\) satisfies \begin{equation}
    \hat{\mathcal{R}}_\pi\vec{\xi}_{\pi, i}=-N^2\xi_{\pi, r, i} \mathbf{Y}_l^m\label{eq:remain}
\end{equation} away from the outer boundary of the acoustic mode cavity,
where \(\xi_{\pi, i}\) are the eigenstates of the modified operator
\(\hat{\mathcal{L}}_\pi\). The matrix elements of
\(\hat{\mathcal{R}}_\pi\) can then be evaluated as the volume integral
\begin{equation}
    R_{\pi\pi, ij} = \left<\xi_{\pi,i}, \hat{\mathcal{R}}_\pi\vec{\xi}_{\pi, j}\right> = -\int \rho N^2\xi_{r, \pi,i}\xi_{r,\pi,j} ~ \mathrm d^3 x.\label{eq:remainpiintegral}
\end{equation} where the spherical harmonic indices \(l, m\) of the
state \(j\) are equal to those of the state \(\pi, i\); the integral
vanishes otherwise. We note that this expression is manifestly
Hermitian. It is also applicable for computing elements of
\(R_{\pi\gamma,ij} = \left<\hat{\mathcal{R}}^\dagger_\pi \xi_{_\pi,i}, \xi_{\gamma,j}\right> = \left<\xi_{\pi,i}, \hat{\mathcal{R}}_\pi \xi_{\gamma_ j}\right>\),
where the state \(j\) is associated with a \(\gamma\) mode rather than a
\(\pi\) mode.

We should in principle be able to do likewise for some \(\gamma\)-mode
remainder operator. However, deriving an exact expression in a similar
manner is less straightforward, as the modification to the oscillation
equations which isolates the \(\gamma\)-mode cavity does not \emph{prima
facie} affect the momentum equation (the first line of \cref{eq:osc} is
instead the perturbed continuity equation with \(\xi_h\) eliminated). We
have not been able to find a corresponding modification to the momentum
equation that yields a manifestly Hermitian expression. For instance, we
might observe that the first line of \cref{eq:osc} can be rewritten in
the form \begin{equation}
    \xi_h = {r \over \Lambda^2}\left[{P_1\over \rho c_s^2} + {\mathrm d \xi_r\over \mathrm d r} + \left({2 \over r} - {g \over c_s^2}\right)\xi_r\right],
\end{equation} with only the first term in the brackets on the right
hand side being suppressed when computing the \(\gamma\)-mode
eigensystem. This suppression can be effected by modifying the
tangential momentum equation (last line of \cref{eq:osc}) to read
\begin{equation}
    -\omega^2 \xi_h = -{1 \over r}\left({P_1 \over \rho} + \Phi_1\right) - \left[{r \omega^2 \over \Lambda^2}{P_1 \over \rho c_s^2}\right],
 \end{equation} where the term in the square brackets does not appear in
the tangential momentum equation of the coupled system. Accordingly the
\(\gamma\) remainder operator might be thought to satisfy
\begin{equation}
    \hat{\mathcal{R}}_\gamma\vec{\xi}_{\gamma, i}= + {r \omega^2_{i, \gamma} \over \Lambda^2} {P_{\gamma, i} \over \rho c_s^2}\mathbf{\Psi}_l^m, \label{eq:gammagamma}
\end{equation} whence \begin{equation}
    R_{\gamma\gamma, ij} = \left<\xi_{\gamma,i}, \hat{\mathcal{R}}_\gamma\vec{\xi}_{\gamma, j}\right> = \int {r \omega^2_{i, \gamma}} \xi_{h,\gamma,i}{P_{\gamma, j} /  c_s^2} ~ \mathrm d^3 x.\label{eq:remaingammaintegral}
\end{equation} Constructions like these do not obviously yield Hermitian
matrix elements, which is problematic in the following sense: since each
of the modified versions of \cref{eq:osc} (for \(\pi\) and \(\gamma\)
modes) yield orthogonal bases with respect to the same inner product as
\(\hat{\mathcal{L}}\), the operators \(\hat{\mathcal{L}}_\pi\) and
\(\hat{\mathcal{L}}_\gamma\) are self-adjoint and Hermitian under that
inner product. It then follows that the remainder operators are also
self-adjoint. Any correct expression for the matrix elements of the
remainder operators must therefore be manifestly Hermitian with respect
to this inner product. Nonetheless, in what follows we will be mostly
concerned with the off-diagonal elements describing the coupling between
\(\pi\)- and \(\gamma\)-modes, which can be expressed entirely in terms
of \(\hat{\mathcal{R}}_\pi\) and the \(\pi\)-mode eigenvalues, and so
this difficulty is not an obstruction to the subsequent analysis. We
will use \cref{eq:remaingammaintegral} to compute only diagonal matrix
elements, and assume that the off-diagonal \(\gamma\gamma\) terms
vanish.

The combined set of basis vectors \(\{\xi_\gamma, \xi_\pi\}\) is not in
general orthonormal, since the \(\pi\) and \(\gamma\)-mode
eigenfunctions are not necessarily orthogonal to each other. However, in
the spirit of \citet{lennard1929electronic}, we can nonetheless express
the general time-dependent state of the linear displacement in terms of
this combined basis as \begin{equation}
    \vec\xi = \sum_i^{N_\pi} c_{\pi,i} \xi_{\pi,i} + \sum_i^{N_\gamma} c_{\gamma,i} \xi_{\gamma,i}.
\end{equation} To find the time evolution of the \(i\)th \(\pi\)-mode
coefficient in particular, we can substitute this into \cref{eq:evol}
and take the inner product against \(\xi_{\pi, i}\) (making use of the
self-adjoint property of all of the operators under consideration) to
obtain \begin{equation}
\begin{aligned}
{\mathrm d^2 \over \mathrm d t^2} c_{\pi, i} + & \sum_j^{N_\gamma} \left<\xi_{\pi, i}, \xi_{\gamma, j}\right> \ddot{c}_{\gamma, j}  =  \\ &-\omega_{\pi, i}^2 c_{\pi, i} + \sum_j^{N_\pi} \left<\xi_{\pi, i}, \mathcal{R}_\pi\xi_{\pi, j}\right> c_{\pi, j} \\&- \sum_j^{N_\gamma} \omega^2_{\pi,i} \left<\xi_{\pi, i}, \xi_{\gamma, j}\right> c_{\gamma, j} + \sum_j^{N_\gamma} \left<\xi_{\gamma, j}, \mathcal{R}_\pi\xi_{\pi, i}\right> c_{\gamma, j}.\label{eq:evolpi}
\end{aligned}
\end{equation} Likewise, the time evolution of the \(\gamma\)-mode
coefficients is given by \begin{equation}
\begin{aligned}
{\mathrm d^2 \over \mathrm d t^2} c_{\gamma, i} &+ \sum_j^{N_\pi} \left<\xi_{\gamma, i}, \xi_{\pi, j}\right> \ddot{c}_{\pi, j} =
\\  &-\omega_{\gamma, i}^2 c_{\gamma, i} + \sum_j^{N_\gamma} \left<\xi_{\gamma, i}, \mathcal{R}_\gamma\xi_{\gamma, j}\right> c_{\gamma, j} \\&- \sum_j^{N_\pi} \omega^2_{\pi,j} \left<\xi_{\gamma, i}, \xi_{\pi, j}\right> c_{\pi, j} + \sum_j^{N_\pi} \left<\xi_{\gamma, i}, \mathcal{R}_\pi\xi_{\pi, j}\right> c_{\pi, j}.\label{eq:evolgamma}
\end{aligned}
\end{equation}

Collecting these coefficients into column vectors \(\mathbf{c}_\pi\) and
\(\mathbf{c}_\gamma\), we can rewrite these expressions in the block
matrix form \begin{equation}
\begin{aligned}
    &{\mathrm d^2 \over \mathrm d t^2}
    \begin{bmatrix}
    \mathbf{c}_\pi \\ \mathbf{c}_\gamma
    \end{bmatrix} \equiv
    \mathbf{L}
    \begin{bmatrix}
    \mathbf{c}_\pi \\ \mathbf{c}_\gamma
    \end{bmatrix} \\ &= 
    \underbrace{\begin{bmatrix}
    \mathbf{I}_{N_\pi} & \mathbf{D}_{\pi\gamma} \\
    \mathbf{D}_{\pi\gamma}^T & \mathbf{I}_{N_\gamma}
    \end{bmatrix}^{-1}}_{\mathbf{G}^{-1}}
    \underbrace{\begin{bmatrix}
        \mathbf{-\Omega}^2_\pi + \mathbf{R}_{\pi\pi} & \mathbf{-\Omega}^2_\pi\mathbf{D}_{\pi\gamma} + \mathbf{R}_{\pi\gamma} \\
        (\mathbf{-\Omega}^2_\pi\mathbf{D}_{\pi\gamma})^T + \mathbf{R}_{\pi\gamma}^T & -\mathbf{\Omega}^2_\gamma + \mathbf{R}_{\gamma\gamma}
        \end{bmatrix}}_{\mathbf{A}}
    \begin{bmatrix}
    \mathbf{c}_\pi \\ \mathbf{c}_\gamma
    \end{bmatrix},
\end{aligned}
\end{equation} where \(\mathbf{I}_n\) is the identity matrix of order
\(n\), the matrix elements of \(\mathbf{R}_{\pi\pi}\) and
\(\mathbf{R}_{\pi\gamma}\) are given by \cref{eq:remainpiintegral},
\(\mathbf{R}_{\gamma\gamma}\) by \cref{eq:remaingammaintegral}, and
\begin{equation}
D_{\pi\gamma, ij} = \int \mathrm d^3 x ~ \rho ~\vec{\xi}_{\pi,i} \cdot \vec{\xi}_{\gamma, j} \label{eq:overlapintegral}
\end{equation} (compare \cref{eq:orthonormal}). Since the matrices
\(\mathbf{A}\) and \(\mathbf{G}\) are both Hermitian, this in principle
defines a generalised Hermitian eigenvalue problem of the form
\begin{equation}
    \mathbf{A}\mathbf{c} = \lambda \mathbf{G}\mathbf{c},
\end{equation} whose eigenvectors are orthogonal with respect to the
inner product \(\mathbf{G}\). However, this makes the subsequent
perturbation analysis extremely unwieldy. Instead, we note that while
the overlap integrals of \cref{eq:overlapintegral} do not, in general,
vanish --- as the \(\pi\) and \(\gamma\) modes are eigenfunctions of
different differential operators --- the \(\pi\)-mode eigenfunctions are
(at least heuristically) oscillatory in the \(\gamma\)-mode evanescent
region, and vice versa, so we expect that \(\max_{i,j}|D_{ij}| \ll 1\),
with these quantities vanishing in the limit of very high or very low
frequencies (where the JWKB approximation holds good). Numerically, we
find this to indeed be the case. We likewise note that the matrix
elements \(\mathbf{R}_{\pi\gamma}\) are also overlap integrals, of a
similar order of smallness relative to the frequency eigenvalues. We
therefore approximate \(\mathbf{L}\) by expanding \(\mathbf{G}^{-1}\) in
series and retaining only first-order terms: \begin{equation}
\begin{aligned}
\mathbf{L} &\sim
    -\begin{bmatrix}
    \mathbf{\Omega}_\pi^2 & 0\\
    0 & \mathbf{\Omega}_\gamma^2
    \end{bmatrix} \\&+ 
    \begin{bmatrix}
    \mathbf{R}_{\pi\pi} & \left(\mathbf{R}_{\pi\gamma}-\mathbf{\Omega}_\pi^2\mathbf{D}_{\pi\gamma} + \mathbf{D}_{\pi\gamma}\mathbf{\Omega}_\gamma^2\right)\\
    \left(\mathbf{R}_{\pi\gamma}-\mathbf{\Omega}_\pi^2\mathbf{D}_{\pi\gamma} + \mathbf{D}_{\pi\gamma}\mathbf{\Omega}_\gamma^2\right)^T & \mathbf{R}_{\gamma\gamma}
    \end{bmatrix}.\label{eq:matrix}
\end{aligned}
\end{equation}

As required, this matrix \(\mathbf{L}\) describes the dynamics of the
coupled \(\pi\) and \(\gamma\) oscillators, whose oscillation
frequencies in isolation are given by the diagonal matrices
\(\mathbf{\Omega}_\pi\) and \(\mathbf{\Omega}_\gamma\). Mixed modes can
be expressed as eigenvectors of this matrix --- i.e.~as linear
combinations of \(\pi\) and \(\gamma\) mode eigenfunctions that
oscillate in phase at the specified frequency eigenvalues, which are in
general distinct from both of the \(\pi\) and \(\gamma\)-mode frequency
eigenvalues.

We now wish to evaluate the eigenvalues of \(\mathbf{L}\), which we
proceed to do perturbatively. Since all of the overlap integrals are
small, and our approximation for \(\mathbf{L}\) is fortuitously
Hermitian, we observe that \cref{eq:matrix} is of the same form as
\cref{eq:perturbation}. For such a decomposition, where \(\mathbf{H}_0\)
has eigenvalues \(E_i\), we recall the standard expression from
perturbation theory for the perturbed eigenvalues in powers of
\(\alpha\) (or \(\mathbf{V}\) as \(\alpha \to 1\)) as
\citep{landau_quantum_1965} \begin{equation}
    E'_i = E_i + \alpha V_{ii} + \alpha^2 \sum_{i \ne j}{|V_{ij}|^2 \over E_i - E_j} + \ldots \label{eq:perturbeig}
\end{equation} To leading order, these are given by the diagonal
elements of the matrix \(\mathbf{L}\), which are dominated, but not
completely specified, by the isolated frequency eigenvalues. Truncating
the series to this order of approximation yields the same result as we
would have obtained with the standard variational approach, treating the
remainder operators as small perturbations to their respective isolated
oscillation equations.

\begin{figure}
\centering
\includegraphics{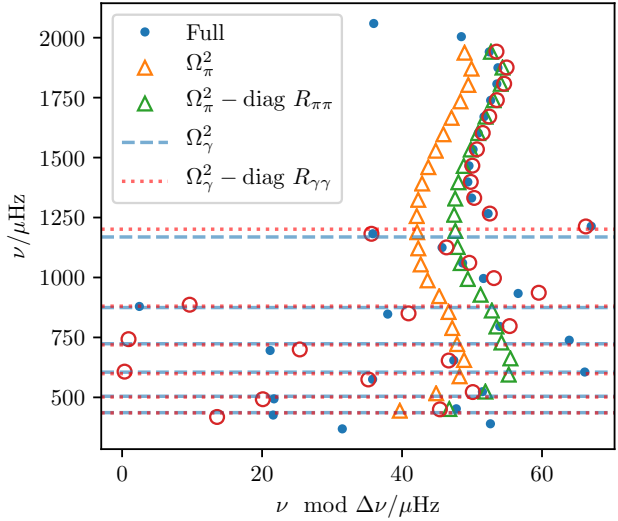}
\caption{Dipole \(\gamma\), \(\pi\), and mixed modes for a \(1M_\odot\)
subgiant model, showing the diagonal contributions from each of the
matrices described in \cref{eq:matrix}. Mixed modes returned from the
full system of equations are shown with the blue points, while the
eigenvalues of the incomplete matrix are shown with red
circles.\label{fig:echelle}}
\end{figure}

To illustrate how these various matrices contribute to the eigenvalues
of the complete coupled system, we show in \cref{fig:echelle} the
predicted isolated and coupled dipole mode frequencies, for a
\(1M_\odot\) subgiant model (described in
\autoref{isolated-eigensystems}). We show the isolated dipole
frequencies of each mode cavity with and without these first-order
corrections. These contributions, though small, cannot be neglected. It
is moreover also apparent that the coupling between the cavities cannot
be derived from first-order considerations; the off-diagonal terms only
enter the series from the second order onward.

In \cref{fig:echelle} we additionally show (with red circles) the
eigenvalues of an incomplete copy of \(\mathbf{L}\) (containing entries
for only the six \(\gamma\) modes shown in the figure). As can be seen,
the accuracy of this incomplete evaluation is increasingly degraded at
low frequencies, both because of the numerical issues we have described
and above, and because the density of missing \(\gamma\)-mode
eigenvalues increases with decreasing frequency --- the sheer number of
\(\gamma\)-mode matrix elements that need to be computed for a complete
result renders this direct approach untenable in the low-frequency
regime even in the absence of implementation-induced numerical
artifacts. Conversely, however, we also see that we obtain good
numerical agreement with the full system of equations in the regime of
individually observed avoided crossings, where \(\gamma\)-modes are
sparse compared to \(\pi\)-modes.

The eigenvectors of \(\mathbf{L}\) specify the relative contributions
from each of the \(\pi\) and \(\gamma\) modes to each mixed mode that
results from the full set of equations. We consider the components of
the \(i\)th mixed mode: \begin{equation}
    \vec{\xi}_i = \sum_j c_{ij}\vec{\xi}_j.
\end{equation} Again, the coefficients \(c_{ij}\) follow from standard
results in perturbation theory: \begin{equation}
\begin{aligned}
    c_{ij} &= \delta_{ij} + \alpha f_{ij} {V_{ij} \over E_i - E_j} \\
    &+ \alpha^2 \left(\sum _k f_{ik}f_{kj} {V_{ik}V_{kj} \over (E_i - E_k)(E_i - E_j)} \right. \\ &\left.- f_{ij}{V_{ii} V_{ij} \over (E_i - E_j)^2} - {1 \over 2} \delta_{ij} \sum_k f_{ik} {V_{ik}^2 \over \left(E_i - E_k\right)^2}\right) + \ldots \label{eq:perturbvec}
\end{aligned}
\end{equation} with \(f_{ij} = 1 - \delta_{ij}\). Generically,
higher-order terms for both the \(i\)th eigenvalues and the eigenvector
components involve increasing powers of resonance/degeneracy factors
\(1/(E_i - E_k)\), which become suppressed for pairs of modes away from
resonance even as \(\alpha \to 1\).

\hypertarget{relation-to-empirical-parameterisation}{%
\subsection{Relation to empirical
parameterisation}\label{relation-to-empirical-parameterisation}}

We contrast this construction with the empirical parameterisation used
elsewhere in the literature
\citep[e.g.~][]{deheuvels_constraints_2011, benomar_masses_2012, benomar_properties_2013},
which is of the generic block form \begin{equation}
    \mathbf{L} = 
    -\begin{bmatrix}
    \mathbf{\Omega_p}^2 & \mathbf{A} \\
    \mathbf{A}^T & \mathbf{\Omega_g}^2 
    \end{bmatrix},\label{eq:2by2}
\end{equation} with \(\mathbf{A}\) an \(N_p \times N_g\) matrix with
constant values along each column, representing \(N_g\) different
coupling constants \(\{\alpha_1 \ldots \alpha_{N_g}\}\);
\(\mathbf{\Omega_p}\) and \(\mathbf{\Omega_g}\) are taken to be diagonal
matrices, related to our quantities as \begin{equation}
\begin{aligned}
    \mathbf{\Omega}_p^2 = \mathbf{\Omega}_\pi^2 - \mathrm{diag}\ \mathbf{R}_{\pi\pi}\\\mathbf{\Omega}^2_g = \mathbf{\Omega}^2_\gamma - \mathrm{diag}\ \mathbf{R}_{\gamma\gamma},\label{eq:diag}
\end{aligned}
\end{equation} while the coupling constants \(\alpha_i\) are explicit
fit parameters. Most practical applications of this parameterisation do
not assume access to the isolated eigenvalues, and therefore supply them
by way of the asymptotic relation (thereby introducing additional,
implicit parameters).

\begin{figure}[hbtp]
    \centering
    \includegraphics[trim=.25cm .25cm 0 .25cm, clip, width=.5\textwidth]{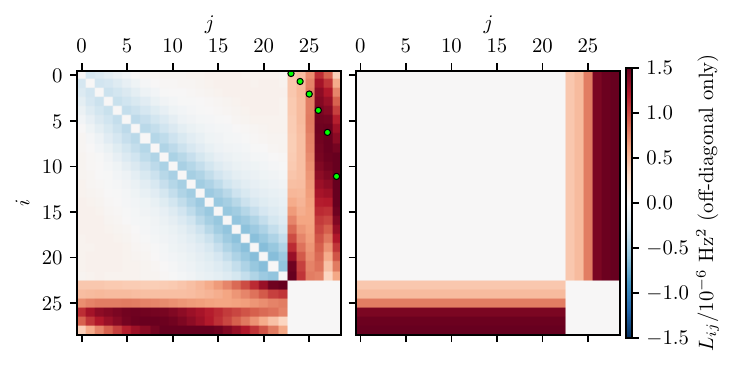}
    \caption{Off-diagonal elements of the matrix $\mathbf{L}$ as computed with respect to a 1$M_\Sun$ subgiant evolutionary model, with modes arranged in order of increasing frequency, and with all $\pi$ modes placed before $\gamma$ modes. In the left panel, we compute these matrix elements using the explicit expressions we have derived in \cref{eq:remainpiintegral,eq:remaingammaintegral,eq:overlapintegral}, while in the right panel we show the corresponding approximate block matrix from the empirical parameterisation given by \cref{eq:2by2}. Green points in the left panel show the indices of the $\pi$-modes (on the vertical axis) closest in frequency to the corresponding $\gamma$-mode (horizontal axis).}
    \label{fig:matrix}
\end{figure}

We compare the off-diagonal elements of \cref{eq:2by2} (right panel)
with those of our explicit construction (left panel) in
\cref{fig:matrix}, for the same set of modes as shown in
\cref{fig:echelle}. The parameters \(\alpha_i\) of the approximate
construction were found by minimising the sum of squared differences
between the ordered eigenvalues of \cref{eq:2by2} and those of our
incomplete matrix.

As noted previously, modes do not couple significantly, irrespective of
the actual value of the corresponding coupling matrix elements, except
where the isolated eigenvalues are close to resonance. To demonstrate
this, we mark out the coupling matrix elements of \(\pi\)- and
\(\gamma\)-mode pairs that are closest to resonance in the left panel of
\cref{fig:matrix}. We see that the best-fitting approximate coupling
parameters in the right panel take values very close to these
on-resonance matrix elements. Moreover, we see that these decrease with
frequency, in line with our expectation that the corresponding overlap
integrals should vanish in the g-mode asymptotic limit
\(\omega^2 \ll S_l^2\).

\hypertarget{relation-to-jwkb-expressions}{%
\subsection{Relation to JWKB
expressions}\label{relation-to-jwkb-expressions}}

Under the JWKB approximation, it is typical to determine eigenvalues by
relating the phase integrals \(\Theta = \int k_r \mathrm d r\) in the g-
and p-mode cavities to each other via a coupling expression of the form
\begin{equation}
    \tan \Theta_p \cot \Theta_g = q, \label{eq:coupling}
\end{equation} where \(q\) is some frequency-dependent coupling
strength, related to the transmission coefficient between the two mode
cavities. Both sides of this expression are understood to be different
functions of frequency, such that mixed-mode eigenvalues are recovered
only at frequencies where this expression holds. In practice, the
frequency dependence of the coupling factor \(q\) is typically neglected
\citep[although see][ for more recent
discussion]{cunha_analytical_2019, pincon_probing_2020, jiang_variations_2020}.

Each of these \(\Theta\) functions yield eigenvalues for the isolated
mode cavities (\(q=0\)) at integer multiples of \(\pi\). The appropriate
constructions in the nonasymptotic regime are of the form
\citep{unno_nonradial_1989, mosser_probing_2012} \begin{equation}
    \begin{aligned}
    \Theta_p = \omega T - \pi\epsilon_{l, p}(\omega) \\
    \Theta_g = {F_l \over \omega} - \pi\epsilon_{l, g}(\omega),\label{eq:angles}
    \end{aligned}
\end{equation} which separately yield the eigenvalue quantisation
conditions for isolated cavities (as in \cref{eq:eigenvalue}). We have
discussed a construction of the buoyancy phase \(\epsilon_g\) above; for
a discussion of the acoustic phase \(\epsilon_p\) see
e.g.~\citet{roxburgh_ratio_2003}. These phases are used, particularly in
the study of red giants, for the computation of a diagnostic quantity
\begin{equation}
    \zeta(\omega) = {I(r_\text{core}) \over  I(R)} \sim \left[1 + {1 \over q}{T \over F_l}\omega^2{\cos^2 \Theta_g \over \cos^2 \Theta_p}\right]^{-1},\label{eq:zeta}
\end{equation} where \begin{equation}
    I(r) = {\int_0^r 4\pi r^2 \rho ~ |\vec{\xi}|^2 ~\mathrm d r \over M \left(\xi_r(R)^2 + \Lambda^2 \xi_h(R)^2\right)}
\end{equation} is the normalisation-independent dimensionless partial
inertia, evaluated up to radius \(r\). This quantity has variously been
used to disentangle the effects of mode bumping from other structurally
or rotationally-induced frequency perturbations
\citep{mosser_period_2015, gehan_core_2018}, or as a
structural/differential rotational diagnostics in its own right
\citep{deheuvels_seismic_2015, deheuvels_near_2017}. By inspection, this
quantity takes values between 0 and 1; for mixed modes of high \(n_g\)
in red giants, it is known to take values close to unity for modes of
predominantly g-like character, and close to zero for modes of
predominantly p-like character. For our purposes, we identify
\(r_\text{core}\) with the inner boundary of the convection zone.

Rather than directly computing \(\zeta\) from our eigenfunctions in this
manner, we first consider the relative contributions to the ratio of
inertiae from a two-term linear combination of the form \begin{equation}
    \vec{\xi} = c_\gamma \vec{\xi}_\gamma + c_\pi \vec{\xi}_\pi. \label{eq:pairwise}
\end{equation} In the limit of both high \(n_g\) and \(n_p\), we recall
that the \(\gamma\)-mode eigenfunction decays rapidly outside of the
convective boundary, while the \(\pi\)-mode eigenfunction does so inside
of it, so to a good approximation the \(\pi\) modes do not contribute
significantly to \(I(r_\text{core})\). Likewise, we expect the overlap
integral matrix elements \(D_{ij}\) to be negligible for the same
reason. We therefore have \begin{equation}
    \zeta = {I(r_\text{core}) \over I(R)} \sim {|c_\gamma|^2 \over |c_\gamma|^2 + |c_\pi|^2}.
\end{equation} By orthonormality (and again neglecting cross-terms
\(D_{ij}\)), the generalisation to a linear combination of many \(\pi\)
and \(\gamma\) modes is immediate: \begin{equation}
    \zeta \sim \left[1 + {\sum_i|c_{\pi, i}|^2 \over \sum_j |c_{\gamma, j}|^2}\right]^{-1}. \label{eq:zeta2}
\end{equation} This expression has the same qualitative properties as
\(\zeta\) --- i.e.~it is close to unity for g-dominated modes and close
to zero for p-dominated modes. While the standard construction of
\cref{eq:zeta} in terms of asymptotic phases relies on JWKB approximants
for the eigenfunctions
\citep[e.g][]{goupil_seismic_2013, deheuvels_seismic_2015},
\cref{eq:zeta2} involves quantities that remain sensible even in the
nonasymptotic regime. We therefore consider \cref{eq:zeta2} to be a
fundamental quantity to which the definitions in \cref{eq:zeta} are
approximations recovered in the JWKB regime. We demonstrate this
explicitly in \autoref{recovery-of-jwkb-expression-involving-zeta}.

\hypertarget{the-coupling-strength-q}{%
\subsection{\texorpdfstring{The coupling strength,
\(q\)}{The coupling strength, q}}\label{the-coupling-strength-q}}

The JWKB coupling strength \(q\) appearing in \cref{eq:coupling} is
given by \begin{equation}
    q = {1 \over 4} \exp \left[- 2 \int_{r_1}^{r_2} \sqrt{- k_r^2}~\mathrm d r\right] \equiv {1 \over 4}w(r_1, r_2),
\end{equation} where \(r_1\) and \(r_2\) are the lower and upper
boundaries of the formal evanescent region between the two mode
cavities, where \(k_r^2 < 0\). In the same construction, the JWKB radial
displacement wavefunction within this evanescent region may be variously
written in the forms \citep[Eqs. 16.47--16.50]{unno_nonradial_1989}
\begin{equation}
\begin{aligned}
    \psi &\sim {A \over \sqrt[4]{- k_r^2}}\left(-{1 \over 2}\sin\Theta_g \exp \left[- \int_{r_1}^r \sqrt{- k_r^2}~\mathrm d r\right] + \cos\Theta_g \exp \left[\int_{r_1}^r \sqrt{- k_r^2}~\mathrm d r\right]\right),\\
    & = {B \over \sqrt[4]{- k_r^2}}\left(-\sin\Theta_p \exp \left[- \int_{r_2}^r \sqrt{- k_r^2}~\mathrm d r\right] + {1 \over 2}\cos\Theta_p \exp \left[\int_{r_2}^r \sqrt{- k_r^2}~\mathrm d r\right]\right),
\end{aligned}
\end{equation} for different choices of constants \(A\) and \(B\). We
identify terms with the isolated radial displacement eigenfunctions in
the following manner: for a two-term linear combination of the form of
\cref{eq:pairwise}, we demand that the component decaying exponentially
as \(r\) increases be identified with \(c_\gamma \xi_\gamma\), while the
component that increases exponentially with \(r\) is to be identified
with \(c_\pi \xi_\pi\). At any radius, the ratio of these two terms
(which we will call \(f(r)\)) must be independent of whether \(A\) and
\(\Theta_g\), or \(B\) and \(\Theta_p\), are used to write the JWKB
wavefunction. As a check of consistency, we should recover
\cref{eq:coupling}. Explicitly: \begin{equation}
\begin{aligned}
    f(r) \equiv {c_\gamma \xi_\gamma \over c_\pi \xi_\pi} \sim -{1 \over 2} \tan \Theta_g w(r_1, r) = -2 \tan \Theta_p w(r_2, r) \\\implies \tan \Theta_p \cot \Theta_g = {1 \over 4} {w(r_1, r) \over w(r_2, r)} = {1 \over 4} w(r_1, r_2) = q.
\end{aligned}
\end{equation} Note that this function \(f\) is regular everywhere in
the domain \([r_1, r_2]\), even though the JWKB wavefunction itself is
singular at the classical turning points (as \(k_r \to 0^-\)). We relate
\(q\) to our quantities via the ratio of \(f\) as evaluated at these
turning points: \begin{equation}
    \begin{aligned}
    {f(r_2) \over f(r_1)} &\sim {4 \tan \Theta_p w(r_2, r_2) \over \tan \Theta_g w(r_1, r_1)} = {4 \tan \Theta_p \over \tan \Theta_g} = 4q \\\implies q &\sim {1 \over 4}{f(r_2) \over f(r_1)} = {1 \over 4} {\xi_\gamma(r_2) \over \xi_\gamma(r_1)} \cdot { \xi_\pi(r_1) \over \xi_\pi(r_2)}.
    \end{aligned}
\end{equation} That is to say, \(q\) is proportional to the product of
the (amplitude) transmission coefficients of the \(\pi\) and \(\gamma\)
waves, considered separately, across the evanescent region.

\hypertarget{applications-to-stellar-modelling}{%
\section{Applications to stellar
modelling}\label{applications-to-stellar-modelling}}

So far, we have concerned ourselves with the theoretical implications of
our construction. In this section we identify and explore ways in which
an explicit isolation of the mode cavities may be applied to modelling
stars against observational seismic constraints, with the ultimate goal
of inferring fundamental stellar parameters.

\hypertarget{pi-modes-for-stellar-modelling}{%
\subsection{\texorpdfstring{\(\pi\)-modes for stellar
modelling}{\textbackslash pi-modes for stellar modelling}}\label{pi-modes-for-stellar-modelling}}

\begin{figure}[htbp]
    \centering
    \includegraphics[trim=0 .75cm 0 .15cm,clip]{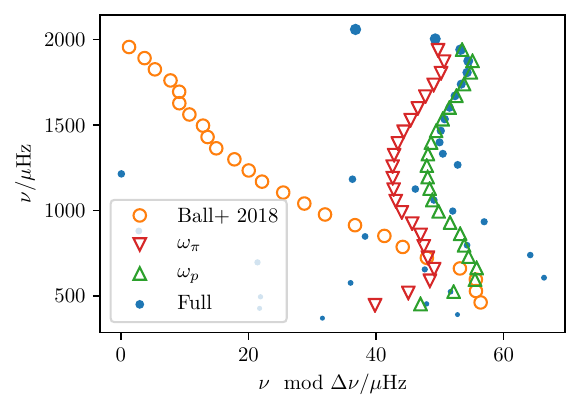}
    \includegraphics[trim=0 0cm 0 .15cm,clip]{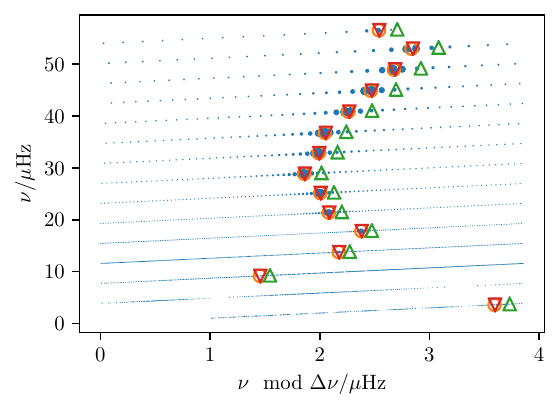}
    \caption{$l=1$ $\pi$-mode eigenvalues as computed with the prescription of \citet[][orange open circles]{ball_surface_2018} and our prescription without (red open triangles) and with (green open triangles) the application of the first-order perturbative correction $R_{\pi\pi}$, for subgiant (upper panel) and first-ascent red giant (lower panel) solar-calibrated 1$M_\Sun$ \texttt{MESA} models. Points are sized inversely to the mode inertia.}
    \label{fig:ball}
\end{figure}

\label{sec:speedup}The prescription of \citet{ball_surface_2018}, while
intended for the same propagation conditions as we are concerned with,
operates by modifying the stellar structure instead of the oscillation
equations. That is to say, where we would set the term \(N^2 \xi_r\) to
zero in the oscillation equations, their prescription does so by
altering \(\Gamma_1\), setting it to \begin{equation}
    \Gamma_{1,\pi} = {\mathrm d \log P \over \mathrm d r} \left/ {\mathrm d \log \rho \over \mathrm d r}\right.
\end{equation} everywhere in the interior radiative zone, which in turn
causes \(N^2\) to vanish. We show the differences between these
approaches in \cref{fig:ball}, for subgiant (upper panel) and
first-ascent red giant (lower panel) solar-calibrated 1\(M_\Sun\)
\texttt{MESA} models. In both cases we compare these results with
mixed-mode eigenvalues computed with respect to the full oscillation
equations and an unmodified stellar model (blue dots).

\begin{figure}
\centering
\includegraphics{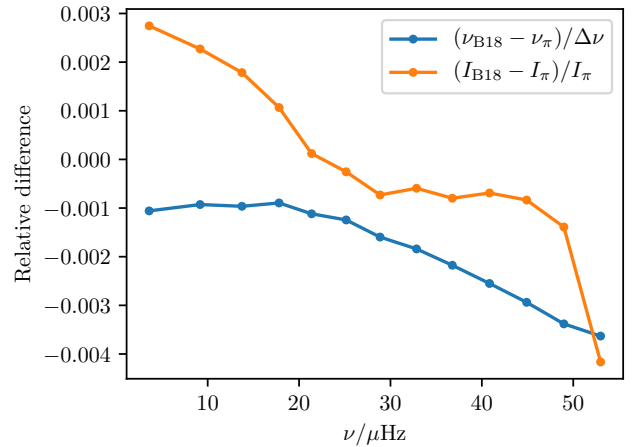}
\caption{Frequency and inertia differences between \(\pi\)-modes
returned from our prescription and those from that of
\citet{ball_surface_2018}. \label{fig:differences}}
\end{figure}

In the regime of isolated avoided crossings, these modifications to
\(\Gamma_1\) yield results that differ significantly from the actual
mixed-mode frequencies, even for modes far from resonance. This is
because the alterations to \(\Gamma_1\) also modify the sound speed
\(c_s^2 = \Gamma_1 P / \rho\), thereby changing the acoustic radius of
the model. This incurs a substantial error in the large frequency
separation \(\Delta\nu\) of the computed frequencies (for which \(1/2T\)
is an asymptotic estimator), which our prescription avoids. By contrast,
our prescription does not modify the stellar structure, but instead
returns \(\pi\)-modes solely from applying pulsation theory; it
correctly recovers the asymptotic behaviour of high-order p-modes.

The prescription of \citet{ball_surface_2018} works better on the red
giant branch --- since the radiative region is very small in physical
extent, the total acoustic radius is not significantly changed. Instead,
modifications to the acoustic mode cavity are confined to a narrow
region of the acoustic radial coordinate, resulting in deviations from
our formulation that are of the form of an acoustic glitch (albeit of
very small amplitude) localised near the inner boundary. We plot these
differences in \cref{fig:differences} (blue curve) --- since these
localised modifications to the model are made fairly close to the centre
of the star, the resulting frequency differences compared to the
\(\pi\)-modes of the unmodified model resemble the effect of some kind
of surface term. The mode inertiae of the modified model are also
changed in a frequency-dependent manner (orange curve). Both of these
effects will necessarily complicate attempts to correct for the true
surface term in actual observational data, e.g.~through
inertia-dependent corrections as in \citet{ball_correction_2014}.

The numerical evaluation of the eigenfrequencies associated with very
evolved red giants is known to be computationally intensive
\citep[e.g.~][]{stello_nonradial_2014}. Leaving aside difficulties
associated with constructing evolutionary models of giant stars in the
first place (which lie beyond the scope of this work), for those stellar
models which we do have, we note that as a star evolves up the red giant
branch, \(\numax\) decreases rapidly compared to the maximum
Brunt-Väisälä frequency in the radiative interior, which instead
increases. The density of \(\gamma\) modes (as given by \(\Delta\Pi\),
\cref{eq:dpi}) therefore increases as the star evolves, and so too does
the density of mixed modes. The majority of these mixed modes are of
very low amplitude (equivalently, have a high mode inertia) near the
surface, as they are primarily g-dominated. Only the most p-dominated
mixed modes, with the lowest inertiae, are typically sufficiently
excited as to be observed. As illustrated in \cref{fig:ball}, the most
p-dominated modes (which have the lowest inertiae) are those that are
closest to resonance with the underlying \emph{uncorrected} \(\pi\)-mode
(i.e.~without the first-order term \(R_{\pi\pi}\)), in keeping with the
dependence of the eigenvector coefficients on the resonance factors in
\cref{eq:perturbvec}. For the purposes of matching observations in these
very evolved stars, it therefore suffices to search only for
\(\pi\)-modes, rather than mixed modes.

Having a high \(n_g\) associated with mixed modes near \(\numax\)
moreover yields a stiff problem in the following sense: let us suppose
that a particular frequency eigenvalue associated with a near-resonance
mixed mode of the form \cref{eq:pairwise} is known in advance. Then the
oscillation equations (expressed via \cref{eq:osc} or
\cref{eq:gyrematrix}) can be cast as an initial value problem, being
integrated outwards from the inner boundary subject to appropriate
initial conditions. Suppose that we integrated this IVP using an
explicit integration scheme; since the \(\gamma\)-mode contribution to
the eigenfunction is highly oscillatory with very short wavelength
(owing to its high order), this suggests that a very small spatial step
size is required for numerical stability (in particular to guarantee
that the \(\gamma\) component decays rapidly outside of the radiative
region). It is therefore the transient component of the stiff system.
Equivalently, when the radial coordinate mesh is refined in the process
of solving the boundary value problem, a very large number of points is
assigned to the radiative zone, in order to assure a sufficiently high
density of mesh points to capture this rapidly oscillatory and
exponential behaviour \citep{jcd_aarhus_2020}.

The ability to compute \(\pi\) modes instead of mixed modes
significantly alleviates both of these computational difficulties; in
\autoref{computational-speedup} we examine a numerical experiment
demonstrating this in more detail.

\hypertarget{grid-based-subgiant-modelling-with-gamma-modes}{%
\subsection{\texorpdfstring{Grid-based subgiant modelling with
\(\gamma\)-modes}{Grid-based subgiant modelling with \textbackslash gamma-modes}}\label{grid-based-subgiant-modelling-with-gamma-modes}}

The frequencies of the lowest-order g-modes evolve rapidly and
monotonically with stellar age as a star evolves off the main sequence
and up the red giant branch. Since they are close to \(\numax\) in
subgiants, these frequencies would place sensitive, surface-insensitive
constraints on stellar ages, were they directly measurable, making these
low-order g-modes particularly valuable. However, these frequencies can
be measured only indirectly via the appearance of the avoided crossing
phenomenon; on the other hand the individual mode frequencies of the
avoided crossing do not evolve monotonically with age, and also evolve
so rapidly (relative to measurement error) as to present difficulties
for grid-based inference of stellar fundamental parameters
\citep{deheuvels_constraints_2011}.

We present a grid-based approach that incorporates age constraints from
avoided crossings. This method requires only that the lowest-order (10
or so) \(\gamma\)-mode frequencies be computed, and is fully
generalisable to cases where multiple avoided crossings are observed. To
illustrate the method, we will examine its application to an actual
subgiant, HD~38529, for which several independent parameter estimates
have been determined from detailed modelling against individual mode
frequencies \citep{ball_robust_2020}. For this purpose we use a grid of
MESA r10398 evolutionary models generated with element diffusion,
without overshoot, with a solar-calibrated mixing-length parameter of
\(1.83\), and with the chemical abundances of
\citet{grevesse_standard_1998}. Models were generated with
\(M/M_\Sun \in [1, 1.6]\) at intervals of \(0.04 M_\Sun\), initial
\(Y \in [0.25, 0.32]\) at intervals of \(0.005\), and initial
\([\textrm{Fe/H}] \in [-0.25, 0.5]\) at intervals of \(0.03\) dex. For
all models we precomputed the frequencies of the first 10
\(\gamma\)-modes in the Cowling approximation, to avoid boundary issues
at high radial order.

\begin{figure}
\centering
\includegraphics{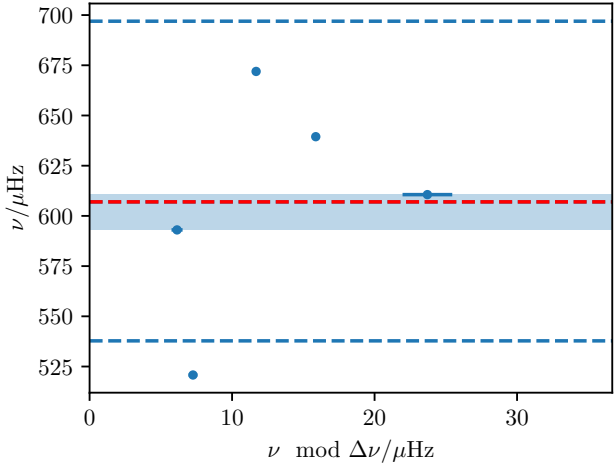}
\caption{Echelle diagram showing measured dipole modes of HD~38529.
Dashed lines show \(\gamma\)-modes computed with respect to the
best-fitting Yale-M model of \citet{ball_robust_2020}, with the
\(n_\gamma = 6\) mode (indicated in red) falling within the search
region defined in the text (shaded interval).\label{fig:38925-echelle}}
\end{figure}

Avoided crossings in this regime are characterised by an extraneous mode
(the \(\gamma\)-mode) disrupting the otherwise regular asymptotic
ordering of the \(\pi\)-modes. Per \cref{eq:perturbeig}, far from
resonance the leading-order effect of mode coupling is to displace the
frequency eigenvalues (relative to the uncoupled \(\pi\)-mode
frequencies) away from the \(\gamma\)-mode, with the two mixed modes
closest to the \(\gamma\)-mode bracketing both it and the on-resonance
\(\pi\)-mode. Since the separation between these is generally less than
asymptotic \(\Delta\nu\), we identify this pair of modes (at frequencies
\(\nu_1, \nu_2\)) as the local minimum of the pairwise frequency
separation between adjacent observed modes (modulo \(\Delta\nu\) to
account for missing modes). We then search for models containing
\(\gamma\)-modes within the interval \([\nu_1, \nu_2]\), as illustrated
for HD~38529 in \cref{fig:38925-echelle}. For the
\(n_\gamma\)\textsuperscript{th} \(\gamma\)-mode at frequency
\(\nu_{n,\gamma}\) associated with a model we define a quantity
\begin{equation}
    g(\nu_{n, \gamma}) = \left\{\begin{array}{cc}
    \nu_{n, \gamma} - \nu_2 & \nu_{n, \gamma} > \nu_2 \\
    0 & \nu_2 \ge \nu_{n, \gamma} > \nu_1 \\
    \nu_{n, \gamma} - \nu_1 & \nu_1 \ge \nu_{n, \gamma}
    \end{array}\right.
\end{equation} from which we construct an associated cost function
\begin{equation}
    C(\nu_{n, \gamma}) = \left[g(\nu_{n, \gamma}) \over (\nu_2 - \nu_1)/2\right]^2.
\end{equation} By construction, \(C(\nu_{n, \gamma})\) is zero when
\(\nu_{n, \gamma}\) lies within our search interval, and grows
quadratically with \(\nu_{n, \gamma}\) outside of it. We therefore
construct an associated weight function \begin{equation}
    w_{n,\gamma} = \exp\left[-{C(\nu_{n, \gamma}) \over 2}\right].
\end{equation} Where multiple avoided crossings are observed, we define
a corresponding number of such search intervals and weights, assigning
consecutively increasing integer values of \(n_\gamma\) to consecutive
avoided crossings in decreasing order of frequency.

Let us consider the case of a single avoided crossing, as seen in
HD~38529. Since the frequencies of all \(\gamma\)-modes increase
monotonically with stellar age, every evolutionary track in the grid
will have some models associated with every \(n_\gamma\) for which
\(w_{n,\gamma} = 0\), and the set of such models forms a series of
nonintersecting hyperplanes, one for each \(n_\gamma\), in the
underlying parameter space.

\begin{figure}
\centering
\includegraphics{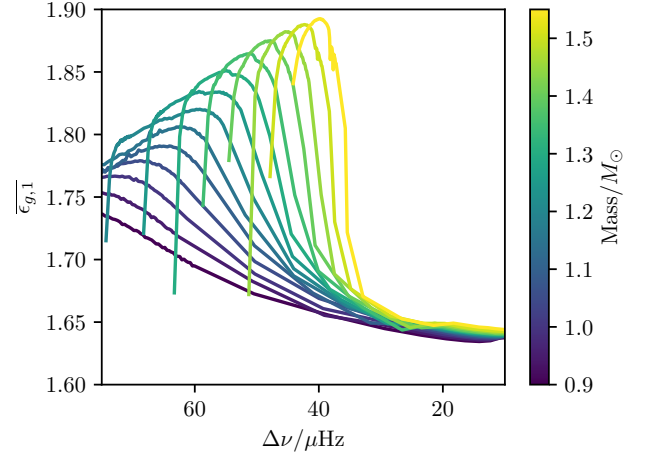}
\caption{Evolution of \(\gamma\)-mode phase \(\epsilon_g\) from
main-sequence turnoff to red giant bump for a sequence of MESA
evolutionary tracks between 0.9 and 1.6 \(M_\odot\). \(\epsilon_g\)
takes values in only a narrow range, permitting \cref{eq:asg} to be used
to identify avoided-crossing radial orders where multiple avoided
crossings can be observed.\label{fig:phase-g}}
\end{figure}

Where multiple avoided crossings are observed, \(\Delta\Pi\) can be
measured and used to estimate \(n_\gamma\) for all avoided crossings for
the star via \cref{eq:asg}, through which any putative identification of
the radial orders can be related to corresponding values of
\(\epsilon_g\). Misidentification of the radial orders is then
equivalent to off-by-one errors in \(\epsilon_g\), which can be ruled
out immediately since \(\epsilon_g\) does not vary significantly (not by
more than a few tenths; see \cref{fig:phase-g}) between main-sequence
turnoff and the red giant bump.

\begin{figure}
\centering
\includegraphics{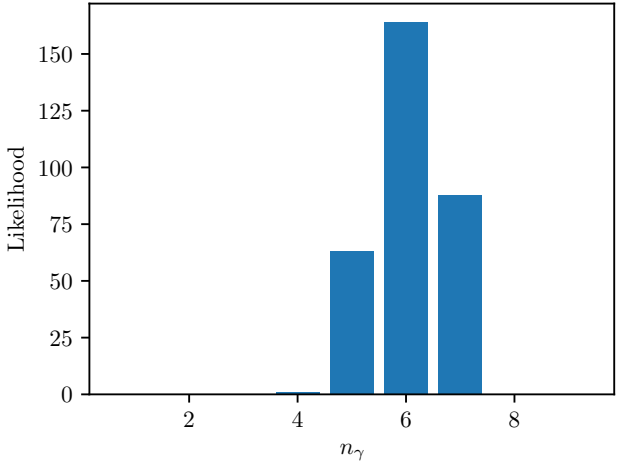}
\caption{Number of models in our model grid with \(w_{n} = 0\) for each
\(n_\gamma\) that also lie within the spectroscopically constrained
\(3\sigma\) region for HD~38529. \label{fig:counts}}
\end{figure}

By contrast, for singly-observed avoided crossings, it is impossible to
identify unambiguously which \(n_\gamma\) is actually responsible for
the avoided crossing in the absence of further information. However, the
hyperplanes we have described above are not all equally favoured by the
spectroscopic observables; we might use e.g.~the number of models with
\(w_n = 0\) that also lie within the spectroscopically constrained
region of parameter space (shown in \cref{fig:counts}) as a proxy for
how likely it is that \(\gamma\)-mode in the actual star is of radial
order \(n_\gamma\). In this case, our model grid suggests
\(n_\gamma = 6\).

Having selected a particular \(n_\gamma\), we then construct an
approximate conditional posterior probability distribution as
\begin{equation}
    p_n \propto w_{n,\gamma} \exp \left[- L_\text{spec} / 2\right],
\end{equation} where \(L_\text{spec} = \sum_i \chi^2_{i, \text{spec}}\).
That is to say, we supplement the ordinary likelihood weights from the
spectroscopic constraints with additional ones from the avoided
crossings, before using them in further analysis (e.g.~to estimate
masses and ages). With multiple avoided crossings, we would take the
product of the weights of all avoided crossings for a given
\(\gamma\)-mode identification.

\begin{figure}
\centering
\includegraphics{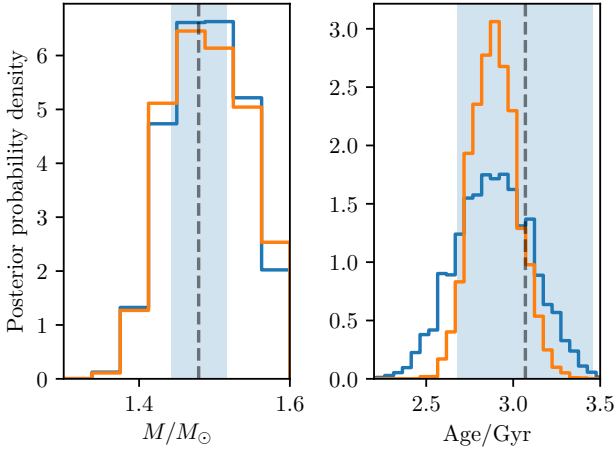}
\caption{Posterior probabilities for stellar mass and age for HD~38529,
with \(n_\gamma = 6\), with (orange curve) and without (blue curve) the
imposition of the avoided crossing constraint. The consensus estimates
from \citet{ball_robust_2020} are shown with the vertical dashed lines,
with the shaded regions indicating consensus uncertainties.
\label{fig:histogram-seis}}
\end{figure}

We show in \cref{fig:histogram-seis} the posterior distributions for
\(n_\gamma = 6\), where we have averaged the posterior distributions
over 100 realisations of Monte-Carlo perturbations of the spectroscopic
constraints from the nominal values as given in
\citet{ball_robust_2020}. Notably, the posterior distribution with the
inclusion of the avoided-crossing weights (orange curve) provides much
strong constraints on the age than the spectroscopic constraints alone
(blue curve). On the other hand, the avoided crossing does not help in
constraining the stellar mass, despite the strong \emph{a priori}
relation between the age at which the avoided crossing is seen
(i.e.~shortly after main-sequence turnoff) and the stellar mass. This
was also the case for the detailed modelling results in
\citet{ball_robust_2020}, where the mass uncertainties returned from
each of the independent detailed modelling efforts was much larger than
would be consistent with the corresponding (very small) age
uncertainties. As they note, this is most likely due to dependences of
the avoided-crossing (i.e.~\(\gamma\)-mode) frequencies on other
compositional or physical parameters.

Finally, while the age uncertainties from these detailed modelling
efforts were small, the corresponding age estimates were largely in
tension with each other. Because of this, the consensus uncertainties
for this detailed modelling work, which includes a contribution from the
internal variance between the different modelling teams, are larger than
even the loose estimates from a coarse grid-based search without any
seismic constraints (right panel of \cref{fig:histogram-seis}). This can
easily be explained by different identifications, by different modelling
teams, of the radial order \(n_\gamma\) of the \(\gamma\)-mode
responsible for the observed avoided crossing: given the paucity of
observed dipole modes, the ambiguity in mode identification that we have
discussed above is also an issue for detailed modelling with individual
frequencies.

\begin{figure}
\centering
\includegraphics{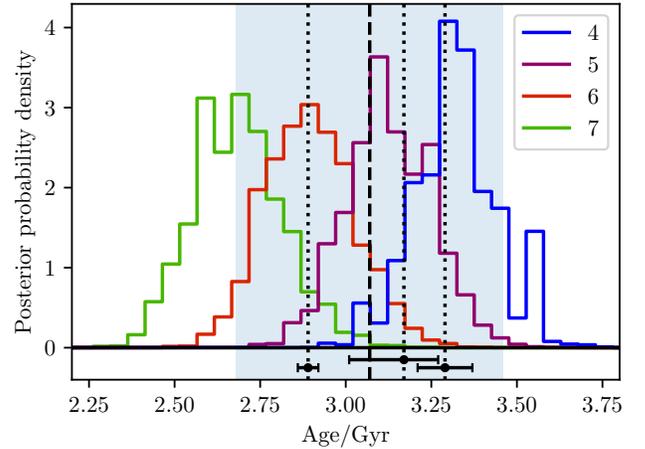}
\caption{Conditional posterior probabilities for stellar age for
HD~38529 associated with different \(n_\gamma\), showing systematic
variation. The consensus result is shown with the vertical dashed line
(with shaded region showing spread between modelling teams), while the
three other dotted lines and black rulers are independent age
constraints from different detailed modelling results, constrained by
individual mode frequencies.\label{fig:histogram-n}}
\end{figure}

We show in \cref{fig:histogram-n} the posterior probability
distributions returned from repeating our Monte-Carlo procedure with
different choices of \(n_\gamma\). The three vertical dotted lines, and
black horizontal rulers, correspond to the nominal ages and
uncertainties returned from three independent detailed modelling
efforts, while the vertical dashed line and light shaded region are the
consensus results and uncertainties. We see that each of these detailed
modelling results lies close to the centre of a conditional posterior
distribution associated with a different \(n_\gamma\). In the absence of
an unambiguous \(\gamma\)-mode identification, which is the case for
HD~38529, the true posterior distribution in the stellar age is best
described as multimodal, likelihood-weighted mixture of these component
distributions. That these detailed modelling results appear to each
sample only one of these component distributions is indicative of more
fundamental methodological issues: for example, optimisation-based
parameter inference is prone to trapping in local optima, which in this
case leads to sampling models with only one value of \(n_\gamma\).
Conversely, were the \(\gamma\)-mode radial order to be specified
\emph{a priori} by some other means, its inclusion as a constraint on
even detailed modelling would most likely alleviate such multimodality.

\hypertarget{discussion-and-conclusion}{%
\section{Discussion and Conclusion}\label{discussion-and-conclusion}}

We have explored different isolation conditions to derive \(\gamma\) and
\(\pi\) modes, which are not uniquely defined and depend on the
propagation structure of the star under consideration. Our chosen
isolation conditions for evolved solar-like oscillators amount to
suppressing terms that would vanish where \(\omega \ll S_l^2\) or
\(\omega \gg N^2\), for the corresponding \(\gamma\) and \(\pi\) mode
cavities, respectively. While these choices are justified based on
asymptotic considerations, the resulting formalism is fully applicable
to all frequency regimes.

The relationship between the isolated and full systems of oscillation
equations is of a form that permits the use of well-established results
from matrix perturbation theory. With respect to this, we have derived
an explicit semi-analytic formulation for the study of various
near-degeneracy phenomena, relating to the coupling between these
isolated cavities. The required matrix elements are expressed as
integrals over the isolated eigenfunctions. Since these constructions do
not rely on the JWKB approximation, they are applicable to buoyancy
waves in subgiants exhibiting avoided crossings, and to acoustic waves
in very evolved red giants, which lie outside the scope of traditional
approaches. Using a numerical implementation based on a general-purpose
pulsation code, we have explored various theoretical consequences and
potential practical applications of this formalism.

Even in cases where the JWKB approach is tenable, access to the
eigenvalues of the isolated mode cavities permits some aspects of the
problem to be simplified. For instance, in many applications where
one-to-one coupling of single \(\pi\)- and \(\gamma\)-mode pairs is
assumed to dominate, the angular quantities in \cref{eq:angles} are
often approximated with \begin{equation}
    \Theta_p(\nu) \sim {\nu - \nu_\pi \over \Delta\nu}, \Theta_g(\nu) \sim \Delta\Pi \left({1 \over \nu} - {1 \over \nu_\gamma}\right),
\end{equation} where the asymptotic relations \cref{eq:asp,eq:asg} are
used to estimate the isolated mode frequencies
\citep[e.g.~][]{mosser_period_2015, gehan_core_2018}. While doing this
is unavoidable where the underlying stellar structure is unknown, this
approach is widely taken even for theoretical studies which do have
access to the underlying stellar structure, resulting in the
introduction of nuisance parameters (\(\Delta\nu, \Delta\Pi\), etc.)
that co-vary with other quantities of interest, thereby complicating the
analysis
\citep[e.g.~][]{benomar_masses_2012, cunha_analytical_2019, jiang_variations_2020}.
We imagine that revisiting these studies with these additional
parameters eliminated might clarify the interpretation of these results.

Finally, we have elucidated a grid-based procedure by which the isolated
\(\gamma\)-mode cavity may be used to constrain global properties of
subgiants undergoing avoided crossings. The process also reveals why age
estimates of sub-giants with only one observed avoided crossing may not
yield the precision suggested by the rapidity of the evolution of any
single \(\gamma\)-mode over an evolutionary track.

\acknowledgements

The authors thank the anonymous referee for the very helpful comments
and suggestions. We thank W. Ball for interesting discussions, and R.
Townsend for technical assistance with GYRE. This research has made use
of the SIMBAD database, operated at CDS, Strasbourg, France; and of
NASA's Astrophysics Data System Bibliographic Services. This work was
partially supported by NASA grant NNX16AI09G to S.B.

\software{NumPy \citep{numpy}, SciPy stack \citep{scipy}, AstroPy \citep{astropy:2013,astropy:2018}, Pandas \citep{mckinney-proc-scipy-2010}, \texttt{MESA} \citep{mesa_paper_1,mesa_paper_2,mesa_paper_4}, \texttt{GYRE} \citep{townsend_gyre_2013}.}

We have made available Python scripts to for various computations in the
Cowling approximation described above, as well as various matrix
elements, at \url{https://gitlab.com/darthoctopus/mesa-tricks}. Our
changes to GYRE (to isolate the \(\pi\) and \(\gamma\) mode cavities,
see \autoref{implementation-details}) can be found in our fork of GYRE
at \url{https://github.com/darthoctopus/gyre}; we have also submitted it
for inclusion upstream.

\appendix

\hypertarget{implementation-details}{%
\section{Implementation details}\label{implementation-details}}

Our modifications to GYRE solve the adiabatic oscillation equations as
expressed in the form \begin{equation}
    x{\mathrm d \over \mathrm d x}
    \begin{bmatrix}
    y_1 \\ y_2 \\ y_3 \\ y_4
    \end{bmatrix}
    =
    \begin{bmatrix}
    {V \over \Gamma_1} - 1 - l_i & {\lambda \over c_1\omega^2} - \alpha_\gamma {V \over \Gamma_1} & {\lambda \over c_1\omega^2} & 0 \\
    c_1 \omega^2 - \alpha_\pi A^* & 3 - U + A^* - l_i & 0 & -1\\
    0 & 0 & 3 - U - l_i & 1 \\
    A^* U & {V \over \Gamma_1} U & \lambda & -(U + l_i - 2)
    \end{bmatrix}
    \begin{bmatrix}
    y_1 \\ y_2 \\ y_3 \\ y_4
    \end{bmatrix}\label{eq:gyrematrix}
\end{equation} where all of these quantities are as given in the GYRE
documentation. For our purposes we need note only that \(x = r / R\),
\(y_1 \propto \xi_r\), \(y_2 \propto P_1\),
\(V / \Gamma_1 = {rg / c_s^2}\), \(A^* = {r N^2 / g}\), and
\(\lambda \to l(l+1)\) for a nonrotating star. Accordingly, the matrix
element (1, 2) corresponds to the coefficient of \(P_1\) in the first
line of \cref{eq:osc}, while the matrix element (2,1) corresponds to
that of \(\xi_r\) in the second line of \cref{eq:osc}.

The isolation of the mode cavities is performed by changing the values
of the newly introduced parameters \(\alpha_\pi\) and \(\alpha_\gamma\),
which are set to 1 by default (in keeping with GYRE's unmodified
behaviour). Setting \(\alpha_\gamma\) to zero yields \(\gamma\) modes,
and setting \(\alpha_\pi\) to zero yields \(\pi\) modes. Additional
allowances have to be made near the inner and outer boundary. For
\(\pi\)-modes in particular, \(\alpha_\pi\) is treated as a function of
radius, and is set to 1 near the surface (defined to be for all \(x\)
larger than some \(x_\text{atm}\), which is supplied as an additional
input parameter) even when \(\pi\)-modes are computed, so as not to
induce a numerical surface term by changing the outer boundary
condition. For \(\gamma\) modes in particular, the eigenfunction
\(\xi_r\) must vanish at both boundaries, for consistency with the
singular nature of the Cowling-approximation Sturm-Liouville problem at
the boundaries.

Additionally, we modify GYRE to estimate the required local density of
the remeshed radial coordinate grid via a dispersion relation of the
form \begin{equation}
    -4 k_r^2 x^2 \sim \gamma = \left(A^* - {V \over \Gamma_1} - U + 4\right)^2 - 4 \left(\alpha_\gamma {V \over \Gamma_1} c_1 \omega^2 - {V \over \Gamma_1}A^* \alpha_\gamma \alpha_\pi - \lambda + {\lambda A^* \over c_1\omega^2} \alpha_\pi\right) =  \left(A^* - {V \over \Gamma_1} - U + 4\right)^2 - 4\left(\alpha_\gamma {V \over \Gamma_1} - {\lambda \over c_1 \omega^2}\right)\left(c_1 \omega^2 - \alpha_\pi A^*\right),\label{eq:gyregrid}
\end{equation} (compare the second term with \cref{eq:dispersion}) where
again the default behaviour is recovered for
\(\alpha_\gamma = \alpha_\pi = 1\).

\hypertarget{recovery-of-jwkb-expression-involving-zeta}{%
\section{\texorpdfstring{Recovery of JWKB expression involving
\(\zeta\)}{Recovery of JWKB expression involving \textbackslash zeta}}\label{recovery-of-jwkb-expression-involving-zeta}}

\hypertarget{zeta-proper}{%
\subsection{\texorpdfstring{\(\zeta\)
Proper}{\textbackslash zeta Proper}}\label{zeta-proper}}

We once again consider a two-term linear combination of the form
\cref{eq:pairwise}. We assume that these modes are close enough to
resonance that we can ignore the contributions from other states, and so
the relevant coupling matrix is the \(2\times2\) matrix of
\cref{eq:2by2} with all entries being scalars. Explicitly, the
eigenvalues are \begin{equation}
    \omega^2_\pm = {\omega_p^2 + \omega_g^2 \over 2} \pm \sqrt{\left(\omega_p^2 - \omega_g^2 \over 2\right)^2 + \alpha^2}, \label{eq:twotermeig}
\end{equation} and the eigenvectors satisfy \begin{equation}
    \begin{bmatrix}
    \omega_p^2 & \alpha \\ \alpha & \omega_g^2
    \end{bmatrix}
    \begin{bmatrix}
    1 \\ u_{\pm}
    \end{bmatrix}
    =\omega_\pm^2
    \begin{bmatrix}
    1 \\ u_{\pm}
    \end{bmatrix},
\end{equation} so we have \begin{equation}
    \begin{aligned}
    \alpha u_\pm &= \omega^2_\pm - \omega_p^2,\\
    {\alpha \over u_\pm} &= \omega^2_\pm - \omega_g^2.
    \end{aligned}\label{eq:eigenvectorcoeff}
\end{equation} Taking the ratio of these, we obtain \begin{equation}
    {1 \over u_\pm^2} = {c_g^2 \over c_p^2} = {\omega_\pm^2 - \omega_g^2 \over \omega_\pm^2 - \omega_p^2}
\end{equation} Without loss of generality, we drop the subscript \(\pm\)
and consider this to be a function of frequency. Since all quantities on
the left-hand-side are positive, this is equal to the ratio of the
absolute values of the numerator and denominator, which we evaluate
separately. For the numerator, we note that \begin{equation}
    |\omega^2 - \omega^2_g| \sim |\delta\omega^2| \sim 2 \omega |\delta\omega| \sim 2\omega^3 \left|\delta\left(1 \over\omega\right)\right| = {2\omega^3 \over F_l}|\Theta_g - n_g \pi|,
 \end{equation} and likewise in the denominator we have \begin{equation}
    |\omega^2 - \omega^2_p| \sim 2\omega|\omega - \omega_p| \sim {2\omega \over T} |\Theta_p - n_p \pi|,
 \end{equation} where \(n_g\) and \(n_p\) are integers. The angular
quantities (as defined in \cref{eq:angles}) are taken to have been
evaluated at the mixed-mode eigenvalues, and so differ only slightly
from integer multiples of \(\pi\). Using a small-angle approximation and
\cref{eq:coupling}, we find \begin{equation}
    {c_g^2 \over c_p^2} \sim {\omega^2 T \over F_l}{|2(\Theta_g - n_g \pi)| \over |2(\Theta_p - n_p \pi)|} \sim {\omega^2 T \over F_l}{|\sin 2(\Theta_g - n_g \pi)| \over |\sin 2(\Theta_p - n_p \pi)|} = {\omega^2 T \over F_l}{|\sin 2\Theta_g| \over |\sin 2\Theta_p|} = {\omega^2 T \over F_l}{|\sin \Theta_g \cos \Theta_g| \over |\sin \Theta_p \cos \Theta_p|} = {1 \over q}{\omega^2 T \over F_l}{|\cos^2 \Theta_g| \over |\cos^2 \Theta_p|}.
\end{equation} Inserting this into \cref{eq:zeta2} yields
\cref{eq:zeta}, as required.

\hypertarget{period-and-frequency-spacings}{%
\subsection{Period and frequency
spacings}\label{period-and-frequency-spacings}}

The frequency and period spacings appearing in \cref{eq:asp,eq:asg} are
some of the easiest seismic observables to relate to evolutionary
properties of stars, and many techniques have been devised to correct
for the effect of mode bumping when measuring them from mixed modes. For
example, \citet{mosser_period_2015} derive a relation between \(\zeta\)
and the local \(\Delta \Pi\) of g-dominated mixed modes in red giants
using the JWKB definition \cref{eq:zeta}, and assuming pairwise mode
coupling, as in \cref{eq:pairwise}. In this section we construct
equivalent statements in the nonasymptotic regime, and derive the
appropriate generalisations to many-mode coupling.

We consider the two isolated mode cavities to yield one ``dense'' and
one ``sparse'' set of eigenvalues. For example, in red giants, we have a
dense series of \(\gamma\) modes (with perturbed, uncoupled frequencies
\(\omega_{g, i}\)) coupling to a sparse series of \(\pi\) modes (at
\(\omega_\pi\)). Since the \(\pi\pi\) and \(\gamma\gamma\) coupling can
be assumed to be weak, we approximate each of the resulting mixed modes
to be a two-term linear combination of the form \cref{eq:pairwise}, with
eigenfrequencies close to \cref{eq:twotermeig}. For
\(\omega_{\gamma,i} < \omega_\pi\), the frequency of the mixed mode in
question is given by \(\omega_-\), and vice versa. If the
\(i\)\textsuperscript{th} uncoupled \(\gamma\)-mode is the closest in
frequency to a given \(\pi\)-mode, the sequence of mixed-mode
eigenvalues goes approximately as \begin{equation}
    \ldots \omega_{-, i-2}, \omega_{-, i-1}, \omega_{-, i}, \omega_{+, i}, \omega_{+, i+1}, \omega_{+, i+2} \ldots
\end{equation} and by assumption the difference between adjacent mixed
modes, \(\delta \omega_\pm\), should tend to the difference between
adjacent uncoupled modes, \(\delta \omega_\gamma\), away from resonance.
We expand this difference
(i.e.~\(\omega^2_{\pm,i+1} -\omega^2_{\pm,i}\)), retaining terms to
first order as \begin{equation}
\begin{aligned}
    \delta \omega_\pm^2 &\sim \delta \omega_\gamma^2\left({1 \over 2} \pm {1 \over 2}{\omega^2_\gamma - \omega^2_\pi \over (\omega^2_+ - \omega^2_-)}\right) \\&= \delta \omega_\gamma^2 {|\omega_\pm^2 - \omega_\pi^2| \over \omega_+^2 - \omega_-^2} = \delta \omega_\gamma^2 {|\omega_\pm^2 - \omega_\pi^2| \over |2 \omega^2_\pm - \omega_\pi^2 - \omega_\gamma^2|},\label{eq:freqdif1}
\end{aligned}
\end{equation} where at each step we have used the relation
\(2 \omega^2_\pm - \omega_\pi^2 - \omega_\gamma^2 = \pm (\omega_+^2 - \omega_-^2)\).
Using \cref{eq:eigenvectorcoeff} we rewrite the above (dropping the
subscript \(\pm\)) as \begin{equation}
    {\delta\omega^2 \over \delta \omega_\gamma^2} \sim {|\alpha u| \over |\alpha \left(u + 1/u\right)|} = \left(1 + u^{-2}\right)^{-1} = \zeta,\label{eq:freqdif2}
\end{equation} which is the result of \citet{mosser_period_2015} if
\(\omega/\omega_\gamma \sim 1\). Completely analogously, for a dense
series of p-modes coupling to a single g-mode, which is typical of
subgiants undergoing avoided crossings, we obtain that \begin{equation}
    {\delta\omega^2 \over \delta \omega_\pi^2} \sim {|\omega^2 - \omega_\gamma^2| \over |2 \omega^2 - \omega_\pi^2 - \omega_\gamma^2|} = {|\alpha / u| \over |\alpha \left(u + 1/u\right)|} = 1-\zeta.\label{eq:freqdif3}
\end{equation}

While these expressions hold near resonance, we would like to consider
cases where the ``sparse'' set of eigenvalues is nonetheless dense
enough that we have one-to-many coupling (one dense mode to many sparse
modes) away from resonance. Indexing the ``dense'' eigenvalues with
\(i\) and the ``sparse'' ones with \(j\), and neglecting coupling
between the dense eigenvalues, we find that (to leading order, per
\cref{eq:perturbeig,eq:perturbvec} with a strictly off-diagonal
perturbation) we can write the corresponding first differences of the
dense eigenvalues as \begin{equation}
\begin{aligned}
    \delta \omega_i^2 &\sim \delta \omega_{0, i}^2 + \sum_{j \ne i} \delta \left(V_{ij}^2 \over \omega_i^2 - \omega_j^2\right)\\
    & \sim \delta\omega^2_{0,i} \left(1 - \sum_{j \ne i} {V_{ij}^2 \over (\omega_i^2 - \omega_j^2)^2}\right)\\
    \implies {\delta \omega^2_i \over \delta\omega^2_{0,i}}& \sim \left(1 - \sum_{j \ne i} c_{ij}^2\right).
\end{aligned}
\end{equation} Here we have assumed that locally the dependence of the
coupling matrix elements on frequency is much weaker than of the
resonance factors. This construction can be applied beyond situations
where \(\delta\omega_i\) is given by asymptotic quantities \(\Delta\nu\)
or \(\Delta\Pi_l\); for example, \citet{deheuvels_near_2017} derive a
similar relation in the case where mode coupling induces an asymmetric
component into the rotational splitting in red giants. We have
demonstrated that these formulations remain approximately applicable in
the nonasymptotic regime, and in particular to subgiant avoided
crossings, subject to a modified definition of \(\zeta\).

\hypertarget{computational-speedup}{%
\section{Computational speedup}\label{computational-speedup}}

\begin{figure}
\centering
\includegraphics{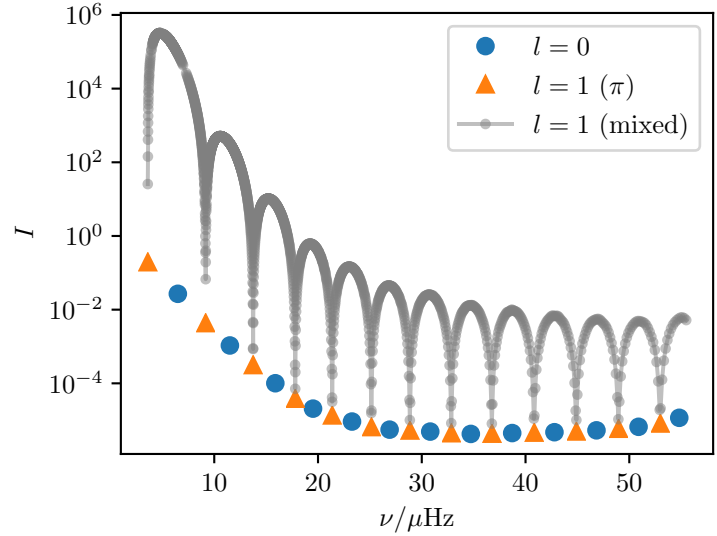}
\caption{Mode inertiae for \(\pi\) and mixed modes, as well as radial
p-modes. The most p-dominated mixed modes are those closest in frequency
to the \(\pi\)-modes, which also can be seen to have the lowest
inertia.\label{fig:inertiae}}
\end{figure}

In \autoref{sec:speedup} we identified two ways in which isolating the
\(\pi\) cavity can speed up the computation of nonradial frequency
eigenvalues from giant stellar models, which we quantify in terms of the
ratio of runtime complexity compared to a direct search for
\(\pi\)-modes:

\begin{itemize}
\tightlist
\item
  The typical search strategy for p-dominated mixed modes involves first
  performing a broad-band search over a range of frequencies, and then
  pruning the results to the most p-dominated mixed modes, as
  characterised by the local minima of the mode inertiae (considered as
  a function of frequency; see \cref{fig:inertiae} for an example).
  Although sufficiently good characterisation of the p-mode asymptotic
  relation may permit this search space to be restricted to a smaller
  number of mixed modes close to the predicted asymptotic p-mode
  frequencies \citep[e.g.~as done in][ for \(l=2\)
  frequencies]{mckeever_helium_2019}, low-order p-modes are known to
  depart severely from the asymptotic relation as stellar models
  approach the tip of the RGB, rendering this approach increasingly
  untenable where it is needed most. By contrast, all \(\pi\) modes
  returned from the computation are guaranteed to be p-dominated,
  requiring no further refinement. The speedup from this goes as
  \(\Delta\nu/\nu^2\Delta \Pi_l\).
\item
  Since the \(\pi\) component of mixed modes constitutes the slow part
  of a stiff system, coarser coordinate meshes can be used for the
  decoupled problem without sacrificing the numerical accuracy of the
  returned eigensystem, than would be possible for the direct
  computation of mixed modes. The speedup factor from this improvement
  goes as \((n_g / n_p)^p\), where the most inefficient \(N\times N\)
  matrix operation in the solution algorithm for the boundary value
  problem has a runtime complexity of \(\mathcal{O}(N^p)\).
\end{itemize}

In practice, we expect the true speedup to be less than this, due to a
combination of low-level systematics (e.g.~time taken for i/o
operations) and other implementation details (e.g.~GYRE only adds mesh
points, and does not take them away).

To better characterise the contributions from each of these factors to
any potential performance gains, we devised an experiment where we
recorded the time taken to perform the following sets of computations
for all models on a MESA evolutionary track:

\begin{enumerate}
\def\labelenumi{\alph{enumi}.}
\tightlist
\item
  Direct computation of \(\pi\)-modes,
\item
  Computation of mixed modes subject to a restricted search strategy,
\item
  Computation of mixed modes subject to a naive search strategy.
\end{enumerate}

\begin{figure}
\centering
\includegraphics{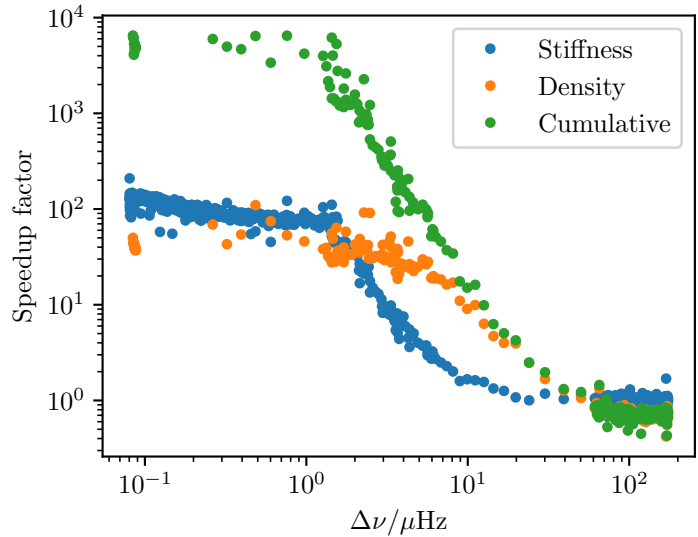}
\caption{Speedup factors (defined as the ratio of runtime complexity of
two different search algorithms). The labels are explained in the main
text. \label{fig:benchmark}}
\end{figure}

The number of mixed modes computed for (b) was chosen to be the same as
(a), so that the speedup from (b) to (a) derives only from decoupling
the transient component of the stiff system. Likewise, the speedup from
(c) and (b) results essentially from only reducing the number of modes
in the search space. We ran this experiment on an Intel Xeon E5-2670 CPU
running at 2.60GHz, using 10 threads (out of 16 cores) for each
computation. In any case, since we are only comparing speedup factors,
we expect these results not to depend on our hardware configuration.
Save for a few models near the tip of the RGB and on the red clump, we
set an upper limit of 1 hr for all computations; this meant that
frequency searches (c) could not be completed for the majority of the
post-bump RGB. Moreover, in all cases we limited ourselves to solving
for at most 1000 eigenvalues; as such, computations for (c) near the tip
of the RGB took much less time than would have a truly exhaustive
search. Nonetheless, as \(\Delta\Pi\) decreases monotonically between
the RGB bump and the tip of the RGB, our results set a lower bound on
the speedup that would be obtained for these circumstances. We show the
results of this experiment in \cref{fig:benchmark}. We see that each of
these reductions in runtime complexity greatly speeds up the search for
eigenvalues for stellar models ascending the red giant branch.

\bibliography{biblio.bib}

\end{document}